\documentclass[preprint]{elsarticle}
\usepackage[utf8]{inputenc}
\usepackage{amsmath}
\usepackage{amssymb}
\usepackage{textcomp}
\usepackage{alltt}
\usepackage{float}
\usepackage{algorithmic}
\usepackage{algorithm}
\usepackage{wrapfig}

\floatstyle{ruled}
\newfloat{program}{thpb}{lop}
\floatname{program}{Algorithm}

\title{Evolutionary-aided negotiation model for bilateral bargaining in Ambient Intelligence domains with complex utility functions}
\author[upv]{Víctor Sánchez-Anguix}
\ead{sanguix@dsic.upv.es}
\author[upv]{Soledad Valero}
\ead{svalero@dsic.upv.es}
\author[upv]{Vicente Julián}
\ead{vinglada@dsic.upv.es}
\author[upv]{Vicente Botti}
\ead{vbotti@dsic.upv.es}
\author[upv]{Ana García-Fornes}
\ead{agarcia@dsic.upv.es}
\address[upv]{Universidad Politécnica de Valencia, Departamento de Sistemas Informáticos y Computación,
Camí de Vera s/n, Valencia, Spain, 46022}

\begin{document}

\begin{abstract}
 Ambient Intelligence aims to offer personalized services and easier ways of interaction between people and systems. Since several users and systems may coexist in these environments, it is quite possible that entities with opposing preferences need to cooperate to reach their respective goals. Automated negotiation is pointed as one of the mechanisms that may provide a solution to this kind of problems. In this article, a multi-issue bilateral bargaining model for Ambient Intelligence domains is presented where it is assumed that agents have computational bounded resources and do not know their opponents' preferences.  The main goal of this work is to provide negotiation models that obtain efficient agreements while maintaining the computational cost low. A niching genetic algorithm is used before the negotiation process to sample one's own utility function (\textit{self-sampling}). During the negotiation process, genetic operators are applied over the opponent's and one's own offers in order to sample new offers that are interesting for both parties. Results show that the proposed model is capable of outperforming similarity heuristics which only sample before the negotiation process and of obtaining similar results to similarity heuristics which have access to all of the possible offers. 
\end{abstract}


\maketitle

\section{Introduction}

Nowadays, the number of computational devices present in our everyday life has grown considerably. The use of technology looks to help us achieve a better quality of life, to make our life easier and more comfortable. However, due to the increasing number of devices, it is necessary that the technology itself adapts to the needs of the user, instead of the human being the one that adapts to technology. In that sense, Ambient Intelligence (AmI) tries to cover that necessity: it looks to offer personalized services and provide users with easier and more efficient ways to communicate and interact with other people and systems \cite{corchado08,weber05}. 

Agent technology has been appointed as a proper technology for the support of AmI solutions \cite{corchado08, fraile09, remagnino05}. In fact, agents show interesting characteristics for AmI environments since they are reactive, proactive and social \cite{Wooldridge01}. Firstly, reactiveness allows agents to change their behavior according to some new conditions in the AmI environment (new users, new services, etc.). Secondly, proactiveness makes it possible for agents to act autonomously according to the user's goals, which results in a smooth and non-intrusive interaction with the AmI user.  And lastly, the agent's social behavior allows several heterogeneous entities to cooperate and offer new complex services to the AmI user.

Over the last few years, researchers in the area of agent technology have shown a growing interest in automated negotiation. Negotiation can be defined as a process in which a joint decision is made by two or more parties. The parties first verbalize contradictory demands and then move towards agreement by a process of concession-making or search for new alternatives \cite{pruitt81}. Therefore, automated negotiation consists in such a joint decision being automatically decided by means of autonomous entities (e.g., agents representing different users). The parties participating in a negotiation process have opposing preferences, thus negotiation can be considered as a conflict resolution mechanism. 

Such conflict circumstances are not alien to AmI applications. For instance, shopping malls may be converted into ubiquitous environments where several vendors offer their products to passing shoppers \cite{keegan08,bajo09}. In many cases, the shoppers know what they want but do not have time to check every shop that offers such products. A possible way of enhancing the customer experience is to automatically negotiate with all of the vendors. A list with the best agreements may be presented to the user through his mobile device. This way, the user does not have to check every possible shop since his mobile device has negotiated with every shop taking into account the user preferences. Nevertheless, there are also benefits for the vendors since automated negotiation allows a more flexible commerce than classic e-commerce. For instance, they may negotiate issues such as price, payment method, discounts, and dispatch dates, which is what often happens in traditional non-electronic commerce. Flexibility in e-commerce may result in client loyalty since the vendor is able to adapt as much as possible to the client preferences. Therefore, automated negotiation is a proper technology for e-commerce-based AmI applications such as shopping malls.

The process of negotiation has been traditionally studied by the field of Game Theory \cite{nash50,rubinstein82}, providing solutions that reach optimal results under different criteria (e.g., Pareto efficiency, Nash Product, etc.). However, such solutions require unbounded computational resources that are not available in most real applications. In such cases, the research area of artificial intelligence (AI) has tried to provide a solution by means of heuristics that achieve results that are as close as possible to the optima \cite{kraus97,jennings01}. Artificial intelligence has traditionally studied multi-issue negotiations where utility functions are represented as a linear combination of the issues involved in the negotiation process \cite{faratin98,faratin00,fatima04a,jonker04}. In linear utility functions, issue values are usually monotonic, so these functions usually have a single global optimum and consequently, the utility function is easy to optimize. Nevertheless, most real world problems  are hardly modeled by linear utility functions since they have a higher degree of complexity than that offered by linear utility functions (e.g. e-commerce \cite{klein02,robu05,ito08} ). Some of the issues in the negotiation setting may present interdependence relationships. Thus, the value of the utility function may be drastically changed by the positive/negative synergy of interdependent issues. The result is that the utility function is no longer linear, and  there may,  therefore,  be several local optima. Optimizing  non-linear utility functions is hard by itself (e.g. it may require non-linear optimizers such as simulated annealing, genetic algorithms, etc.), as is learning opponent preferences and looking for good agreements. Utility functions that have the trait of being non-linear are usually known in the literature as complex utility functions. 

Over the last few years, there has been an effort to research negotiation strategies that are capable of working with such complex utility functions where issues may have relationships of interdependence. Works in these complex domains have focused on negotiation strategies that require a mediator \cite{klein02,ito08,marsa09,marsa09b}, or non-mediated strategies that are devised for very specific utility functions \cite{robu05}. However, non-mediated strategies are more interesting from the point of view of AmI environments due to the fact that users enter and leave the system in an extremely dynamic way. Thus, it may be difficult to find a trusted mediator for every possible user. Although non-mediated strategies are more interesting from the point of view of different domains, there has been a lack of work in non-mediated strategies for complex utility functions. The work of Lai et al. \cite{lai08} presents a non-mediated strategy for general utility functions, which obviously includes complex utility functions. The strategy is based on the calculation of current iso-utility curves and a similarity heuristic that sends offers from the current iso-utility curve that are the most similar to the last offers received from the opponent. However, the entire calculation of the iso-utility curve may require an exhaustive exploration of the utility function, which may not be tractable in the case of a large number of issues. Furthermore, if the exploration of one's own utility function is not performed in an intelligent way, the result may be that most of the offers sampled are of no use for the negotiation process since they might not interest the opponent. Mechanisms that sample as few offers as possible are needed, especially for environments where devices may have limited computational resources as is the case with AmI environments.

In this work, a non-mediated bilateral multi-issue negotiation model for AmI environments is presented. Its main goal is to optimize  the computational resources while maintaining a good performance in the negotiation process. The proposed model is inspired by the seminal work of Lai et al. \cite{lai08}. The three main differences between this present work and the work of Lai et al. are: (i) The present approach assumes that it is not possible to exhaustively search the utility function. Before the negotiation process starts, each agent samples its own utility function by means of a niching genetic algorithm (GA) \cite{holland75,mahfoud95}. The effect of this sampling is that offers obtained are highly fit and significantly different;(ii) A few additional samples are obtained during the negotiation process by means of genetic operators that are applied over received offers and one's own offers. The heuristic behind this sampling is that offers obtained by genetic operators have genetic material from one's own agent and the opponent's offers. Thus, these new offers may be interesting for both parties. (iii) Genetic operators act as a learning mechanism that implicitly guides the offer sampling and selection of which offers must be sent to the opponent. 
 
Results show that the proposed work outperforms similarity heuristics that are able to sample the same number of offers before the negotiation process starts. Additionally, it is also shown how the proposed strategy is capable of achieving similar results to those of similarity heuristics that sample the entire utility function with far fewer samples. This result is accomplished due to the learning mechanism provided by genetic algorithms.

This paper is organized as follows: section 2 describes an example of application where automated negotiation and ambient intelligence can be combined in order to offer a useful service for the user; section 3 describes the negotiation model, explaining the chosen protocol and the new negotiation strategy in detail. In Section 4, the experimental setting and the results obtained are discussed. In Section \ref{related} related work is discussed. Finally, the conclusions and future lines of work are explained in Section 6.

\section{An Example of Automated Negotiation and Ambient Intelligence Synergy: Product Fairs}
In this section we introduce an example of application where automated negotiation may be used along with well-known AmI techonologies in order to provide a profitable service for users. The example is focused on product fairs. Fairs are public events where manufacturers/sellers/producers exhibit their products to a wide range of consumers who go from small consumers to big retailers. At this kind of events there are usually a large number of exhibitors and products. Therefore, it is extremely difficult to explore the whole fair or find interesting deals for one's interests. It is also difficult for sellers to attract interesting clients. Thus, both consumers and sellers would be benefited by a tool which allows them to attract/search prospective deals quickly. 

At this point automated negotiation in an AmI environment may come in handy. Let us suppose the following scenario at a fair: each vendor has been assigned a booth where he attends to clients. As well as setting up the typical equipment, a hardware device with Bluetooth wireless communication is provided (e.g. a PC). An agent, which can be downloaded and configured by the vendor prior to the fair event day, is installed in this hardware device, and it complies with rules and communication protocols established by fair organizers. These agents should be provided with information regarding its owner's preferences by means of user modeling methods such as questionnaires, past experiences, and so forth. 

Additionally, consumers are allowed to download an agent to their mobile devices prior to the fair event. The only requirement for the mobile device is Bluetooth wireless capabilities. The consumer's agent follows the communication protocols established by the fair organizers and can be configured similarly to the vendor's agent. More specifically, the agent may ask what its owner would be interested in buying and general questions about the preferences regarding the possible negotiation attributes.

When consumers and vendors enter the fair, they should start the execution of their respective agents. Each consumer's agent offers a negotiation service which can be invoked by vendors' agents. Whenever this service is invoked by a vendor agent, a negotiation process starts between the vendor agent and the consumer agent. The negotiation process continues until a deal has been found or one of the parties has decided to leave. If the deal is considered as interesting by both parties (i.e. utility of the deal higher than a certain threshold or reservation utility) and the deal is among the best ones for the consumer in that specific area (determined by which vendors can be reached by Bluetooth in that space point), the consumer agent and the vendor agent notify their respective owners regarding the possible deal. However, deals discovered by this automatic process are not to be considered as binding but as recommendations. If the deal is considered as interesting enough by the consumer, it may result in the consumer approaching the vendor's booth. At that point, both parties may decide to renegotiate or polish the deal which has been found by their agents.

Since Bluetooth technology has coverage limitations, the service can usually only be discovered by vendor agents that are nearby. Therefore, negotiation processes help consumer and vendor agents to find prospective deals as consumers walk around the fair. More specifically, it allows consumers to save physical time by recommending them the vendors that seem more suitable for their needs in the area. That way, they only approach vendors in an area who may have an interesting deal for them. Indirectly, it also helps vendors since it attracts consumers with high probabilities of buying their goods instead of losing time with clients with whom the possibilities of making a deal are very low. 

\section{Negotiation Model}
\label{sec:negmod}
Negotiation models are composed of a negotiation protocol and a negotiation strategy. On the one hand, the negotiation protocol defines the communication rules to be followed by the agents that participate in the negotiation process. More specifically, it states at which moments the different agents are allowed to send messages and which kind of messages the agents are allowed to send. For instance, the Rubinstein alternating protocol specifies \cite{rubi90} that agents are allowed to  send one offer in alternating turns. Basically, the negotiation protocol acts as a mechanism for the coordination and regulation of the agents that take part in the negotiation process.

On the other hand, the negotiation strategy defines the different decisions that the agent will make at each step of the negotiation process. It includes the opponent's offers acceptance rule, the selection of which offers are to be sent to the opponent, the concession strategy, the decision of whether the agent should continue in the negotiation process or not, and so forth. Therefore, the negotiation strategy includes all the decision-making mechanisms that are involved in the negotiation process.

The negotiation protocol used can be categorized as an alternating protocol for bilateral bargaining \cite{rubi90}. More specifically, the protocol used is the \textit{k-alternating protocol} proposed by Lai et al. \cite{lai08}. The proposed negotiation strategy belongs to the family of negotiation strategies that use a similarity heuristic in order to propose new offers to the opponent \cite{faratin00,lai08}.

\subsection{Negotiation Protocol}
As mentioned above, the negotiation protocol belongs to the family of alternating protocols for bilateral bargaining. In this kind of protocols, two different agents negotiate without the need of a mediator. As previously stated, non-mediated strategies are more adequate for AmI applications since users enter and leave the AmI system in a very dynamic way. Thus, it may not be feasible to find a trusted mediator for every possible pair of agents. Furthermore, in some AmI domains such as shopping malls, where there are different competing vendors and lots of potential users, it is difficult to determine who will mediate the negotiation process.

The protocol used is the \textit{k-alternating protocol} proposed by Lai et al. \cite{lai08}. This protocol is composed of several rounds where the agents exchange offers in an alternating way. One of the agents, called the \textit{initiator}, is responsible for starting the current round. He can accept one of the previous offers received from the opponent in the last round, exit from the negotiation process, or send up to $k$ different offers to the opponent agent. Once the \textit{initiator} has performed one of the possible actions, the opponent agent is able to accept one of the offers he has just received, exit from the negotiation process or propose up to $k$ different offers to the initiator. Then, the round ends and a new round is initiated by the \textit{initiator} agent. The negotiation process ends when one of the agents accepts an offer (the negotiation succeeded) or one of the agents decides to abandon the negotiation (the negotiation failed).

Some of the properties of the \textit{k-alternating protocol} proposed by Lai et al. are: 
\begin{itemize}
 \item The protocol is adequate for situations where both agents are equal in power (e.g. none of them has the monopoly over a resource).
 \item Each agent is capable of sending up to $k$ different offers, making it more probable that one of the proposed offers satisfies the requirements of the opponent agent.
 \item Since $k$ different offers are proposed in each agent's turn more information about opponent preferences can be inferred, increasing the chances of finding a \textit{win-win} situation. This may produce faster agreements, which is inherently interesting for every domain but particularly for AmI domains since it may reduce the number of messages exchanged and thus the bandwidth consumption.  
\end{itemize}

\begin{figure}
 \centering
 \includegraphics[width=\linewidth]{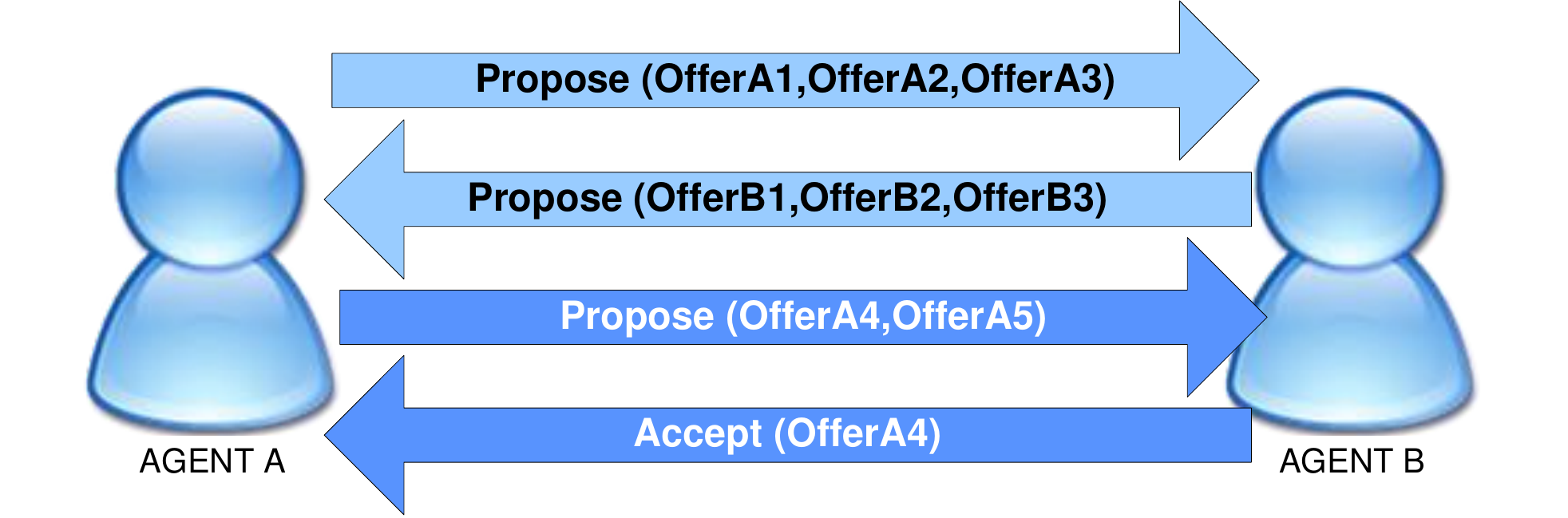}
 \caption{An example of two agents in the k-alternating protocol proposed by Lai et al. \cite{lai08}}
 \label{fig:protocol}
\end{figure}

An example of two agents negotiating with a \textit{3-alternating protocol} ($k=3$) can be observed in Figure \ref{fig:protocol}. Agent A is the initiator of the negotiation round, whereas Agent B is the responding agent. The first round starts with 3 offers proposed by the initiator. Once the offers reach Agent B, he decides whether he should accept one of them or not. Since the 3 offers are not interesting for Agent B, he decides to counteroffer 3 different offers. When the 3 first offers from Agent B reach Agent A, the second round starts. Due to the fact that none of the offers proposed by Agent B are of interest to the initiator, he decides to send 2 offers. The 2 offers from the initiator reach Agent B, who analyzes the offers in order to determine whether they are interesting. Since he found OfferA4 to be interesting, he decides to accept it and the protocol thus ends with an agreement.

\subsection{Negotiation Strategy}
 The proposed negotiation strategy can be classified within the group of strategies that use similarity heuristics to propose new offers to the opponent \cite{faratin00,lai08}. The proposal complements some of the benefits introduced in the inspiring work of Lai et al. \cite{lai08}, making it especially interesting for AmI environments. The goal is to optimize  the computational resources while maintaining a good performance in the negotiation process. The main traits of the proposed model are twofold. Firstly, it is not necessary to sample the entire utility function. Secondly, the proposed strategy provides an implicit learning mechanism that guides the offer sampling and which of the offers sampled are to be sent to the opponent. 

The different decision-making mechanisms of the negotiation strategy can be grouped according to the period during which they are applied: pre-negotiation and negotiation. The former group of decision making is applied before the negotiation process starts. Basically, since utility functions are complex and it is not feasible to completely explore them, each agent samples its own utility function by means of a niching GA (\textit{self-sampling}). 

The latter group of mechanisms is applied during the negotiation process. It includes the acceptance criteria for opponent offers, the concession strategy, the \textit{evolutionary sampling}, and the selection of which offers are to be sent. The most remarkable part is introduced with \textit{evolutionary sampling}: genetic operators are carried out over received offers and one's own offers in order to sample new offers that may be of interest to both parties.  \textit{Evolutionary sampling} acts as an implicit learning mechanism of the opponent's preferences. The result of  evolutionary sampling may be used afterwards when the offers to be sent to the opponent are selected. A brief outline of the proposed strategy can be observed in Algorithm \ref{fig:brief}. A more detailed outline of the strategy used before the negotiation process and during the negotiation process can be observed in Algorithms \ref{fig:niching} and \ref{code}.

\begin{program}
 \Large \textbf{Negotiation Strategy}\\\\
  \underline{\large \textbf{Pre-negotiation}}\\
       \normalsize \textbf{1.}Self-sampling\\\\
  \underline{\large \textbf{Negotiation Process}}\\
         \textbf{2.}Receive opponent offer(s) if there are any offers\\
         \textbf{3.}Acceptance criteria: accept an offer and end the negotiation, or reject all of them and continue the negotiation process\\
         \textbf{4.}Concession strategy\\
         \textbf{5.}Evolutionary sampling\\
         \textbf{6.}Select which offers to send\\
         \textbf{7.}Send offer(s) and go to step 2\\

\caption{A brief outline of the negotiation strategy}
\label{fig:brief}
\end{program}

\subsubsection{Pre-negotiation: Self-sampling}
When an agent uses complex utility functions to represent its preferences it may be difficult to find own offers with good utility. If the number of issues is not very large the complete sampling of the utility function may be feasible. However, when the number of issues is large, this complete sampling may be an extremely expensive process. For instance, a complete sampling of a negotiation domain formed by 10 integer issues from 0 to 9 requires sampling $10^{10}$ offers. The cost associated to this sampling can be exorbitant, especially if agent preferences change with a frequency that is greater than the time invested in the sampling. Furthermore, this sampling is unacceptable for AmI domains. Not only does it take too much computational time and power, but it would also need too much storage for the limited devices usually found in these domains. The sampling process can be reduced by skipping offers that are of very low quality for the agent (i.e., offers with utility equal to zero).

A possible solution to this problem is to use mechanisms that enable to sample good offers for the negotiation process and skip those of low quality. Due to the highly non-linear nature of complex utility functions, non-linear optimizers are required for this task. The main goal is to sample a set of different offers that have good utility and are significantly different, because these offers may point to different regions of the negotiation space where a good deal may be found for the agent.

In this work, a genetic algorithm (GA) was used to solve this problem. GA's are general search and optimization mechanisms based on the Darwinian selection process for species \cite{goldberg89,holland75}. Genetic operators such as crossover, mutation, and selection are employed in order to find near-optimal solutions for the required problem. Nevertheless, the problem posed by classic GA's is that the entire population converges to one optimal solution. As already stated, different interesting offers for the negotiation process need to be explored. Niching methods are introduced to confront problems of this kind \cite{mahfoud95,mengshoel08}. These methods look to converge to multiple, highly fit, and significantly different solutions.
 
A possible family of niching methods for GA's is the crowding approach \cite{mengshoel08}. Crowding methods achieve the desired result by introducing local competition among similar individuals. One advantage of crowding methods is that they do not require parameters beyond the classic GA's. Euclidean distance is usually used to assess the similarity among individuals. Probabilistic Crowding ($P_{C}$) and Deterministic Crowding ($D_{C}$) \cite{mengshoel08} are two of the most popular crowding methods. They only require a special selection rule with respect to classic GA's. Both rules are employed to select a winner given $n$ different individuals. On the one hand, $D_{C}$ selects the individual that has the highest fitness value, resulting in an elitist selection strategy. On the other hand, $P_{C}$ allows lower fitness value individuals to be selected as winners with a certain probability. This probability is usually proportional to the fitness of each individual. $P_{C}$ behavior is more exploratory than $D_{C}$. In both cases, the niching effect is achieved by applying either of the two rules to those individuals that are similar.  Each parent is usually paired with one of its children in such a way that the sum of the distances between pair elements is minimal. For each pair, one of the two crowding rules is employed to determine which individuals will form the next generation. $D_{C}$ and $P_{C}$ can be observed in more detail in Equations \ref{dcr} and \ref{pcr}, respectively.

\begin{equation}
 D_{c}(s_{1},s_{2})= \left\{ 
\begin{array}{l l}
  s_{1}& f(s_{1})>f(s_{2})\\
  s_{2}& f(s_{1})<f(s_{2})\\
  s_{1}\vee s_{2} & \mbox{other} \end{array} \right.
\label{dcr}
\end{equation}
\begin{eqnarray}
 P_{c}(s_{1},s_{2})= \left\{ 
\begin{array}{l l}
  s_{1}& f(s_{1})>f(s_{2})\; \wedge\;rand\leq p_{1}\\
  s_{2}& f(s_{1})>f(s_{2})\; \wedge\;rand > p_{2}\\
  s_{2}& f(s_{1})<f(s_{2})\; \wedge\;rand \leq p_{2}\\
  s_{1}& f(s_{2})<f(s_{1})\; \wedge \;rand > p_{1} \\
  s_{1}\vee s_{2}& other \end{array} \right.
\label{pcr}
\end{eqnarray}
\begin{eqnarray*}
 \mbox{with   } p_{i}=\frac{f(s_{i})}{f(s_{i})+f(s_{i'})}
\end{eqnarray*}
where $rand \in [0,1]$, $f(.)$ is the fitness function, $s_{1}$ and $s_{2}$ are two solutions, and $p_{1}$ and $p_{2}$ are the probability of acceptance of both solutions given the pair $(s_{1},s_{2})$.

The designed mechanism uses a GA that employs crowding methods to find significantly different good offers. This GA is individually executed by the agent before the negotiation process begins. The chromosomes of this GA represent possible offers in the negotiation process, whereas the fitness function used is one's own utility function. A portfolio with $D_{C}$ and $P_{C}$ is used. The population has a fixed number of individuals and the whole population is selected to form part of the genetic operator pool. Pairs of parents are selected randomly and multi-point crossover or mutation operators are applied over them. In both cases, the result is two children. Each parent is paired with the child that is more similar to it according to Euclidean distance. $P_{C}$ or $D_{C}$ is applied to each of the pairs according to an established probability $p\sb{dc}$ and $1-p\sb{dc}$ respectively. Those individuals that are selected as winners by the crowding replace the current generation. The stop criterion was set to a specific number of generations.  At the end of the process, the whole population should have converged to different good offers that are to be used by the negotiation process as an approximation to the real utility function of the agent. This population, called $P$,  is used as an input for the negotiation process. A more detailed outline of the proposed GA can be observed in Algorithm \ref{fig:niching}.

\begin{program}
 
 \begin{alltt}
\small
  \(P:\) Explored preferences, good quality offers
  \(D\sb{c}:\) Deterministic crowding rule
  \(P\sb{c}:\) Probabilistic crowding rule
  \(p\sb{cr}:\) Probability of crossover operator
  \(p\sb{dc}:\) Probability of DC
  \(n:\) Current number of generations
  \(n\sb{max}:\) Maximum number of generations
  \(pair\sb{i}:\) Pair of solutions

  Initialize \(P\)
  \(n=0\)  
  
  Do
      \(n=n+1\)
      shuffle \(P\)
      \(P\sb{aux}=\emptyset\)
      \(i=1\)
      While \(i\leq |P|-1\)
          \(p\sb{1}=P\sb{i}\)
          \(p\sb{2}=P\sb{i+1}\)
          If Random() \(\leq p\sb{cr}\)
              \((c\sb{1},c\sb{2})=crossover(p\sb{1},p\sb{2})\) 
          Else
              \(c\sb{1}=mutate(p\sb{1})\)
              \(c\sb{2}=mutate(p\sb{2})\)
          EndIf
          \((pair\sb{1},pair\sb{2})= \underset{\begin{array}{l}\scriptstyle p\sb{i}\neq p\sb{j}\\\scriptstyle c\sb{k}\neq c\sb{l}\end{array}}{\operatorname{argmin}} ||p\sb{i}-c\sb{k}||+||p\sb{j}-c\sb{l}|| \)

          If Random() \(\leq p\sb{dc}\)
              \(Add(P\sb{aux},D\sb{c}(pair\sb{1}))\)
              \(Add(P\sb{aux},D\sb{c}(pair\sb{2}))\)
          Else
              \(Add(P\sb{aux},P\sb{c}(pair\sb{1}))\)
              \(Add(P\sb{aux},P\sb{c}(pair\sb{2}))\)
          EndIf
          \(i=i+2\)
      EndWhile
      \(P=P\sb{aux}\)
  While \(n\leq n\sb{max}\)
  Return \(P\)
 \end{alltt}

 \caption{Pre-negotiation: Genetic algorithm with niching mechanism. Its goal is to sample the agent utility function}
 \label{fig:niching}
\end{program}

\hyphenation{A-gents}
\subsubsection{Negotiation: Concession strategy}
A concession strategy determines which utility the agent will try to achieve at each negotiation step. The agent usually proposes offers that have a utility equal or above the utility level defined by its concession strategy at a specific negotiation round. In this work, we assume a time-dependent strategy, where the utility required by each agent depends on the remaining negotiation time. This kind of concession strategies are adequate for environments such as AmI, where time is a limitation (e.g., limited power devices, goods that loose their value as time passes, real-time environments, etc.). Some examples of concession strategies are \textit{sit-and-wait} \cite{an08} (no concession until the deadline, e.g. one of the agents has monopoly), linear (same concession rate at each step), boulware \cite{faratin98,pruitt81} (no concession until the last rounds, where it quickly concedes to the reservation value), and conceder \cite{faratin98,raiffa82} (at the start, it quickly concedes to the reservation value).

One of the traits of similarity-based strategies is that they are usually independent of the underlying concession strategy. However, this work assumes an environment where agents have similar market power (similar concession rate), and similar computational resources (similar deadlines). Thus, a linear concession strategy is assumed. 

In each negotiation round, the agents concede according to their strategy until a private deadline is reached. The minimum utility that an agent $a$ demands for a negotiation round $t$ can be formalized as follows:

\begin{equation}
 U_{a}(t)=1-(1-RU_{a})(\frac{t}{T_{a}})\pm \delta
\end{equation}
where $U_{a}(t)$ is the minimum demanded utility level for agent $a$ at negotiation round $t$, $RU_{a}$ is the reservation utility, $\delta$ is a small value that allows to accept/select offers which are relatively close, and $T_{a}$ is the private deadline of the agent. 

\subsubsection{Negotiation: Acceptance criteria}

The acceptance criteria for an agent usually depend on its concession strategy. Normally, an opponent offer is accepted if it provides a utility that is equal or greater than the demanded utility for the next negotiation round. Consequently, given the set of offers $X_{b\rightarrow a}^{t}$ received by agent $a$ from agent $b$ at instant $t$, the acceptance criteria for agent $a$ can be formalized as depicted in the following expression:
\begin{eqnarray}
 Accept_{a}^{t}(X_{b\rightarrow a}^{t})= \left\{ 
\begin{array}{l l}
  \mbox{accept} & V_{a}(x_{b\rightarrow a}^{t,best})\geq {U_{a}(t+1)}\\
 & \\
  \mbox{reject} & otherwise
\end{array} \right.
\end{eqnarray}
where $Accept_{a}^{t}(X_{b\rightarrow a}^{t})$ is the offer acceptance function, $V_{a}(x)$ valuates the utility of an offer, $x_{b\rightarrow a}^{t,best}$ is the best offer received from the opponent at negotiation round $t$, and ${U_{a}(t+1)}$ is the utility demanded for the next negotiation round.

\subsubsection{Negotiation: Evolutionary sampling}

 One of the keys of the proposed strategy is \textit{evolutionary sampling}. This provides an implicit mechanism for learning opponent preferences and making an intelligent sampling. Basically, it is based in the application of some genetic operators to offers received from the opponent in the last negotiation round and one's own good offers from $P$. The idea behind the \textit{evolutionary sampling} is that offers generated by this method have genetic material from the opponent and one's own agent. Therefore, these offers may yield a greater probability of being accepted by the opponent that offers that have been sampled in a blind way. The new offers are added to a special population called $P_{evo}$ which contains offers that have been generated by genetic operators. 

Let us consider $X_{b\rightarrow a}^{t}=[x_{b\rightarrow a}^{t,1},x_{b\rightarrow a}^{t,2},...,x_{b\rightarrow a}^{t,k}]$, which is the set of offers sent by agent $b$ to agent $a$ at negotiation round $t$, and $U(t)$ the current desired utility to generate offers at negotiation round $t$.  For each offer $x_{b\rightarrow a}^{t,i}$, a total of $M$ offers are selected from the current iso-utility curve $IC_{P}$ (offers with a utility equal to $U(t)$) defined in the population $P$. These $M$ offers minimize the expression:

\begin{equation}
 \underset{\begin{array}{l}C\in IC_{P} \\ |C|=M \end{array}}{\operatorname{argmin}} \sum_{j=1}^{M} ||x_{b\rightarrow a}^{t,i}-c_{j}||
\end{equation}
where $C$ is the set of $M$ different offers, and $||x_{b\rightarrow a}^{t,i}-c_{j}||$ is the Euclidean distance between one of the offers in $C$ and the offer received from the opponent. Thus, these $M$ offers are the ones most similar to $x_{b\rightarrow a}^{t,i}$ from iso-utility curve in $P$ and they will be involved in the evolutionary process. Offers are selected from the current iso-utility curve since offers with much greater utility may generate new offers with a utility that is no longer useful in the negotiation process (e.g. a utility greater than the current utility), and offers with lower utility may produce new offers that are not to be used until the last rounds of the negotiation process. Furthermore, the $M$ selected offers are the most similar since applying crossover operators over offers that are too different may disrupt the quality of the solution for both agents (the resulting offer is too far from both agents' offers).

\begin{figure}
\begin{center}
\includegraphics[width=226px,height=207px]{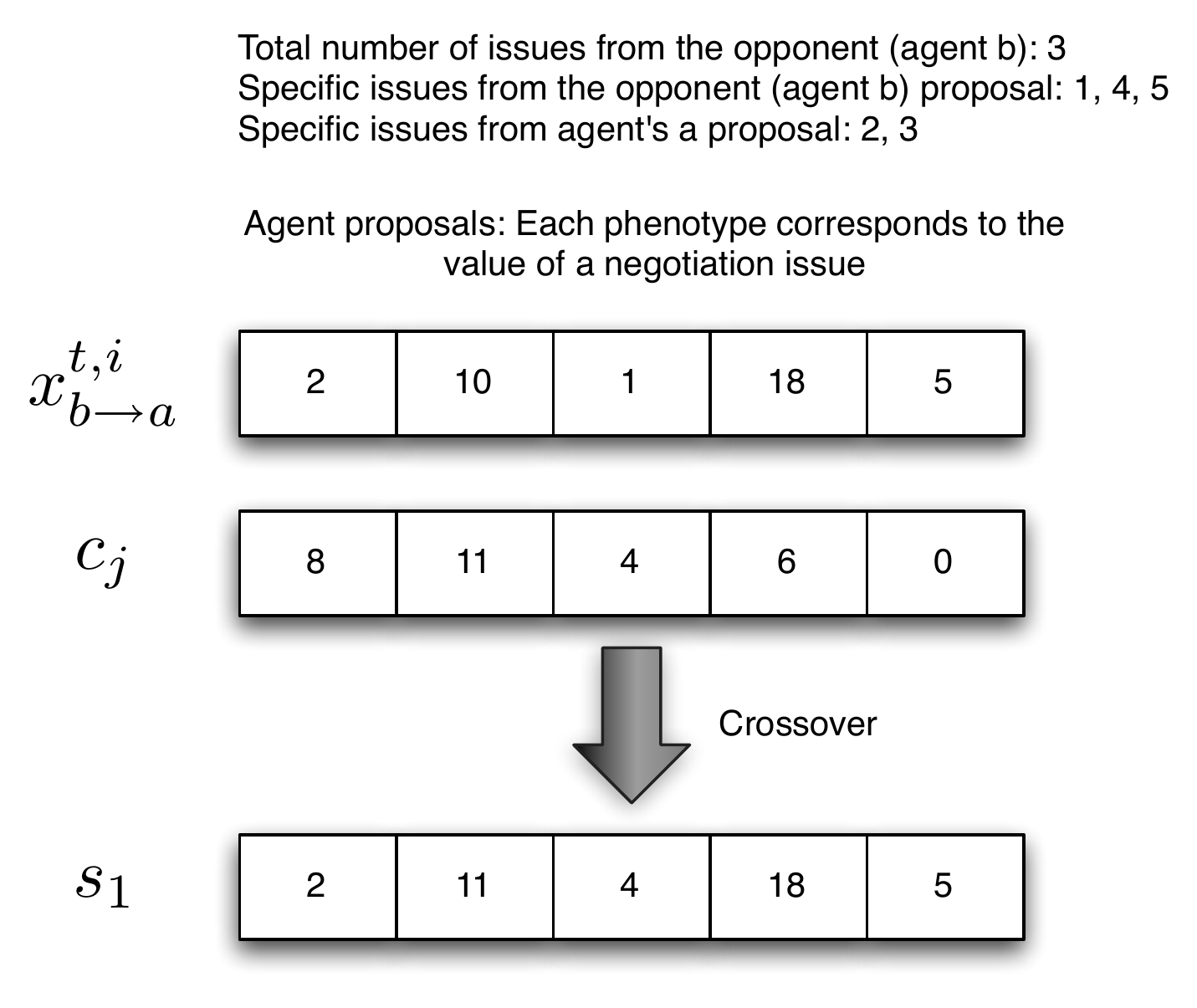}
\caption{An example of a crossover operation}
\label{fig:cross}
\end{center}
\end{figure}

Once the $M$ closest offers have been selected, a total of $n_{cross}$ crossover operations are performed for each pair $(x_{b\rightarrow a}^{t,i},c_{j})$, where $c_{j} \in C$. The crossover operator takes two parents and generates one child. More specifically, the number of issues that come from $x_{b\rightarrow a}^{t,i}$ is chosen randomly from $1$ and $N-1$, with $N$ being the number of issues. The rest of the issues come from $c_{j}$. Which particular issues come from each parent is also decided randomly. This way, each agent's preferences are taken into account in a statistically equal manner. Each child is added to a special pool, called $P_{evo}$,  that contains new offers sampled during the different \textit{evolutionary sampling} phases. An example of a crossover operation can be observed in Figure \ref{fig:cross}.

A total of $n_{mut}$ mutation operations are carried out for each generated child by crossover operations. The mutation operator changes issue values randomly, according to a certain probability of mutating individual issues ($p_{attr}$). When $p_{attr}$ is low, mutated offers are close to the original offer, so the effect is the exploration of the neighborhood of the offer. The operator is applied $n_{mut}$ times to each child that is produced by crossover operations and to the original offers from the opponent. Mutation also generates new children that are added to the special pool $P_{evo}$

Note that no offer from $P_{evo}$ is discarded even though their utility may be considered too low for the current negotiation round. The reason for this mechanism is that offers that are not currently acceptable may be interesting in future negotiation rounds due to the concession strategy. Furthermore, since they have genetic material from the opponent's offers, they are more likely to be accepted.

As can be observed in Algorithm \ref{code}, if the negotiation process lasts $n_{round}$ rounds, the Evolutionary Sampling will have explored a total number of offers that is equal to:
\begin{eqnarray*}
 Samples_{evo}=n_{round}*( (k*M*n_{cross}) + (k*M*n_{cross})*n_{mut} + k*n_{mut} ) \\
 =n_{round}*k*( M*n_{cross}*(1+n_{mut})+n_{mut}     )
\end{eqnarray*}
Then, the number of offers sampled during the negotiation process depends on the number of rounds that the negotiation lasts, $k$, $M$, and the number of genetic operators that are performed per offer selected from the iso-utility curve.

\subsubsection{Negotiation: Select which offers to send}

The next step in specifying the negotiation strategy consists of defining the mechanism to propose new offers. In this case, it is necessary to devise a mechanism that is capable of proposing up to $k$ different offers to the opponent and taking into account the preferences of the opponent. The applied heuristic takes into account the $k$ offers received from the opponent and the offers in $P$ and $P_{evo}$. 

In order to select these offers, $k$ offers from the current iso-utility curve are sent. More specifically, two different iso-utility curves are calculated. The first one is the iso-utility curve calculated using offers in $P$, called $IC_{P}$. The second one is the iso-utility curve calculated using offers in $P_{evo}$, called $IC_{E}$. Basically, the first iso-utility curve has offers that were generated during the \textit{self-sampling} (only taking into account one's own preferences), whereas the second iso-utility curve only has offers that were generated in the \textit{evolutionary sampling} (they may take into account both agents' preferences).  The negotiation strategy defines a proportion of $p_{pevo}$ offers to come from $IC_{E}$. The rest of the offers come from $IC_{P}$.

The offers selected from $IC_{E}$ are those that minimize the distance to any offer received from the opponent in the previous negotiation round. This selection may be formalized as:

\begin{equation}
\underset{\begin{array}{l}\scriptstyle C\subset IC_{E}\\ \scriptstyle|C|=p_{pevo}*k \end{array}}{\operatorname{argmin}} \left( \sum_{j=1}^{C} \underset{x \in X_{b\rightarrow a}^{t}}{\operatorname{min}}||c_{j}-x|| \right)
\end{equation}

On the other hand, offers are also selected from $IC_{P}$. The total number of offers corresponds to a proportion that is equal to $1-p_{pevo}$. In this case, offers that are the closest to any offer received from the opponent in the previous negotiation round are selected. This selection can be formalized as:

\begin{equation}
\underset{\begin{array}{l}\scriptstyle D\subset IC_{P}\\ \scriptstyle|D|=(1-p_{pevo})*k \end{array}}{\operatorname{argmin}} \left( \sum_{j=1}^{D} \underset{x \in X_{b\rightarrow a}^{t}}{\operatorname{min}}||d_{j}-x|| \right)
\end{equation}

The parameter $p_{pevo}$ determines the degree of relevance of the new offers sampled during the \textit{evolutionary sampling} with respect to the offers sampled before the negotiation process. When $p_{pevo}=0$, the strategy ignores the results that come from $P_{evo}$. Consequently, only offers that were sampled in the pre-negotiation phase (\textit{self-sampling}) are sent to the opponent.  In this particular case, the strategy is equivalent to a negotiation strategy that only samples before the negotiation process and does not take into account the opponent's preferences. In contrast, when $p_{pevo}=1$, the offers sampled during the \textit{evolutionary sampling} are the only ones taken into account. In any case, $p_{pevo}$ is a parameter to be adjusted.

\begin{program}
 \begin{alltt}
\small
\(P\): Offers from \textit{self-sampling}         \(P\sb{new}\): Offers from \textit{evolutionary sampling}
\(k\): Number of offers of the protocol  \(M\): Number of selected offers
\(n\sb{cross}\): Number of times to crossover  \(n\sb{mut}\): Number of times to mutate
\(p\sb{pnew}\): Proportion of offers from \(P\sb{new}\)

Receive \(X\sb{b\rightarrow a}\sp{t}\)

If \(V\sb{a}(x\sb{b\rightarrow a}\sp{t,best})\geq U\sb{a}(t+1)  \) then Accept

Update current utility t=t+1

/*Evolutionary sampling*/

For each \(x\sb{b\rightarrow a}\sp{t,i} \) in \(X\sb{b\rightarrow a}\sp{t}\)
     \(C= \underset{ \begin{array}{l}\scriptstyle C\subset IC\sb{P}\\ \scriptstyle|C|=M \end{array}}{\operatorname{argmin}} \sum\sb{j=1}\sp{M} ||x\sb{b\rightarrow a}\sp{t,i}-c\sb{j}|| \)

    For each \( c\sb{j} \) in \(C\)
        Repeat \(n\sb{cross}\) times
           s1=Crossover(\(x\sb{b\rightarrow a}\sp{t,i},c\sb{j} \))
           If s1 \(\nsubseteq\) \(P\sb{new}\) then Add(\(P\sb{new},\)s1)
           Repeat \(n\sb{mut}\) times
               s2=Mutate(s1)
               If s2 \(\nsubseteq\) \(P\sb{new}\) then Add(\(P\sb{new},\)s2)
           EndRepeat
        EndRepeat
    EndFor
    Repeat \(n\sb{mut}\) times
        s1=Mutate(\(x\sb{b\rightarrow a}\sp{t,i}\))
        If s1 \(\nsubseteq\) \(P\sb{new}\) then Add(\(P\sb{new},\)s1)
    EndRepeat
EndFor

/*Select which offers to send*/
\(k\sb{1}=p\sb{pnew}*k\)
\(X\sb{1}= \underset{\begin{array}{l}\scriptstyle C\subset IC\sb{E}\\ \scriptstyle|C|=k\sb{1} \end{array}}{\operatorname{argmin}}  \sum\sb{j=1}\sp{C} \underset{x \in X\sb{b\rightarrow a}\sp{t}}{\operatorname{min}}||c\sb{j}-x||  \)


\(k\sb{2}=(1-p\sb{pnew})*k\)
\(X\sb{2}= \underset{\begin{array}{l}\scriptstyle D\subset IC\sb{P}\\ \scriptstyle|D|=k\sb{2} \end{array}}{\operatorname{argmin}}  \sum\sb{j=1}\sp{D} \underset{x \in X\sb{b\rightarrow a}\sp{t}}{\operatorname{min}}||d\sb{j}-x||  \)
 
 
\(X\sb{a\rightarrow b}\sp{t+1}= X\sb{1}\cup X\sb{2}\)

Send \(X\sb{a\rightarrow b}\sp{t+1} \)
\end{alltt}
\caption{Negotiation strategy during the negotiation process}
\label{code}
\end{program}

\subsubsection{Case Study}

We prepared a very simple case based on the product fair example. To be more specific, it depicts a purchase in a furniture fair where one buyer is interested in buying chairs and tables from a seller. It should be pointed out that the goal of this case study is not to test the performance of the model, which will be thoroughly studied in Section \label{exp}, but to show show a trace of the negotiation model from the point of view of one of the agents. In this case, we will focus on the buyer. 

We used the weighted constraint model proposed by Ito et al. \cite{ito08} to represent the utility functions of the buyer and the seller. The weighted constraint model was introduced as a complex utility function to model agent preferences. Let us consider a negotiation model where the number of issues is $N$, $s_{i}$ represents the i-th issue, each issue has a domain $s_{i} \in [0,X]$ that sets its maximum and minimum value, and $\vec{s}=(s_{1},s_{2},...,s_{N})$ represents a particular offer. These settings make up an N-dimensional space for the utility function.

In the weighted constraint model, a constraint $c_{l}$ represents a specific region of the space. Any point of the space enclosed in that region is said to satisfy the constraint $c_{l}$. Basically, the term \textit{constraint} represents an interdependence relationship among the negotiation issues. Each constraint $c_{l}$ has a certain value $v(c_{l},\vec{s})$ that is added to the utility of $\vec{s}$ when the constraint is satisfied by the point $\vec{s}$. For instance, a constraint defined as $c_{l}= (1\leq s_{1} \leq 10 \wedge 3\leq s_{2}\leq 4)$  and $v(c_{l},\vec{s})=10$ would hold a utility of 10 for the point (2,3) of the space.

A utility function in the weighted constraint model is formed by $l$ constraints whose values are summed up whenever the constraints are satisfied. The utility of a point $\vec{s}$ given $l$ constraints can be defined as:
\begin{equation}
 U(\vec{s})=\sum_{c_{l}\in L} v(c_{l},\vec{s})
\end{equation}
where $\vec{s}$ is the point/offer, $c_{l}$ is a constraint, $L$ is the set of constraints, and $v(c_{l},\vec{s})$ is the value of the constraint if it is satisfied (0 otherwise).

As stated in \cite{ito08}, although the expression seems linear, it produces a non-linear utility space due to the interdependence among the issues represented by the constraints. Furthermore, the utility function may generate spaces with several local maxima, which makes the problem highly non-linear and very difficult to optimize. Additionally, the agents do not have any knowledge about the possible constraints of the opponent, thus making the problem of negotiation still more difficult.

This negotiation case consists of 3 different attributes: price (P) [0-9], chair model (CM) [0-9], and table model (TM) [0-9]. Next, we introduce the utility functions we employed to represent the preferences of both consumer and seller:
\begin{center}
	\scriptsize
	\begin{tabular}{|l|l|}
	\hline
 	Buyer Utility Function & Seller Utility Function \\
	\hline
	($v_{1}=100$) $(0\leq P \leq 1)$ & ($v_{1}=80$) $(8\leq P \leq 9) $\\
	($v_{2}=50$) $(2\leq P \leq 4) $ & ($v_{2}=60$) $(6\leq P \leq 7)$  \\
	($v_{3}=25$) $(5\leq P \leq 7) $ & ($v_{3}=45$) $(4\leq P \leq 5)$ \\
($v_{4}=30$) $(0\leq CM \leq 3)  \wedge (0\leq TM \leq 3)$ & ($v_{4}=20$) $(1\leq P \leq 3)$   \\
($v_{5}=10$) $(0\leq CM \leq 3)  \wedge (6\leq TM \leq 9)$ & ($v_{5}=15$) $(1\leq CM \leq 2) $\\
($v_{6}=50$) $(0\leq CM \leq 3)  \wedge (5\leq TM \leq 6)$ & ($v_{6}=10$) $(0\leq CM \leq 1) $\\
($v_{7}=30$) $(4\leq CM \leq 6)  \wedge (0\leq TM \leq 3)$ & ($v_{7}=10$) $(2\leq CM \leq 5)  $ \\
($v_{8}=20$) $(4\leq CM \leq 5)  \wedge (4\leq TM \leq 5)$ & ($v_{8}=5$) $(5\leq CM \leq 9) $ \\
($v_{9}=10$) $(4\leq CM \leq 5)  \wedge (8\leq TM \leq 9)$ & ($v_{9}=20$) $(8\leq CM \leq 9) $ \\
($v_{10}=50$) $(7\leq CM \leq 9)  \wedge (2\leq TM \leq 4)$ & ($v_{10}=60$) $(0\leq TM \leq 1)  $\\
($v_{11}=20$) $(7\leq CM \leq 9)  \wedge (6\leq TM \leq 8)$ & ($v_{11}=30$) $(1\leq TM \leq 4)  $ \\
                                                            & ($v_{12}=5$) $(4\leq TM \leq 6) $\\
                                                            & ($v_{13}=20$) $(6\leq TM\leq 9) $\\
                                                            & ($v_{14}=10$) $(8\leq TM\leq 9) $\\
	\hline
	\end{tabular}
\end{center}

The consumer shows attribute interdependences relating the two types of furniture (e.g. some pairs of models fit better than other pairs). In the case of the seller, no interdependences are found but he may present preferences regarding which models to sell (e.g. some of them need to be manufactured; some models only have a few units, etc.).

As for the parameters of the self-sampling phase, they were set to $|P|=16$, $n_{max}=100$, $p_{dc}=80\%$ and $p_{cr}=80\%$. The rest of parameters of the negotiation model were set to $\delta=0.05$, $k=2$, $T=10$, $p_{pevo}=100\%$, $n_{cross}=2$, $n_{mut}=2$, and $M=2$. 

The next table shows the 16 offers found by the self-sampling process carried out by the buyer. It depicts the value for each attribute and the utility of the offer. In this case the utility has been scaled to [0,1] for the sake of simplicity.

\begin{center}
 \scriptsize
 \begin{tabular}{|l  l|}
 \hline
  \multicolumn{2}{|c|}{P=Self-sampling results for the buyer} \\
 \hline
  $(u=1.00)$ 1 1 6 & $(u=0.81)$ 1 3 0 \\
  $(u=1.00)$ 0 1 6 & $(u=0.81)$ 1 5 3 \\
  $(u=0.93)$ 1 7 3 & $(u=0.62)$ 0 2 4 \\
  $(u=0.93)$ 1 7 4 & $(u=0.62)$ 1 9 1 \\
  $(u=0.93)$ 1 9 2 &  \\
  $(u=0.93)$ 1 2 5 & \\
  $(u=0.93)$ 1 8 3 & \\
  $(u=0.93)$ 1 8 4 & \\
  $(u=0.93)$ 0 7 3 & \\
  $(u=0.93)$ 1 9 3 & \\
  $(u=0.93)$ 0 1 5 & \\
  $(u=0.81)$ 1 5 0 & \\
 \hline
 \end{tabular} 
\end{center}


\paragraph{Round 1 $U_{s}(1)=1-0.95$ $U_{b}(1)=1-0.95$}
Once the self-sampling phase has ended, the negotiation process starts with the buyer acting as initiator. Since there are no opponent offers to value, evolutionary sampling is skipped and the agent directly proposes offers to the opponent. Due to the fact that no evolutionary sampling has been carried out, $P_{evo}$ is empty and only the iso-utility curve which can be calculated comes from P. X=(1 1 6) and Y=(0 1 6) are randomly selected since there is no opponent offer to compare with. The opponent rejects the offers since they yield a utility of 0.35 and 0.25 respectively. The opponent makes a counteroffer which contains W=(8 1 1) and Z=(9 1 1). Both of them are rejected since their utilities (0.18 for both of them) are lower than $U_{s}(2)=0.85$.
\paragraph{Round 2 $U_{s}(2)=0.95-0.85$ $U_{b}(2)=0.95-0.85$} 
Two offers have been received from the opponent. Thus, the evolutionary sampling phase is carried out. The iso-utility curve from P ($U_{s}(2)=0.95-0.85$) is shown in the following tables. It shows the offers and the euclidean distance to W and Z. For both W and Z, the $M=2$ offers which are more similar are selected. These offers selected from the iso-utility curve become one of the parents for the genetic operations, which are also shown in the following tables. For the sake of simplicity, genetic operations which produced children that were already in $P_{evo}$ are not included (nor are they stored more than once). All of the offers generated during this phase are added to $P_{evo}$.

\hspace{-1.5cm}\begin{minipage}{0.27\linewidth}
 \scriptsize
 \begin{tabular}{|l | l | l|}
\hline
 \multicolumn{3}{|c|}{Iso-utility curve (P)} \\
 \hline
 Offer & d(W) & d(Z) \\
 \hline
 1 2 5 & 0.90 & 1.00 \\
 0 1 5 & 0.99 & 1.09 \\
 1 7 3 & 1.04 & 1.13 \\
 1 7 4 & 1.07 & 1.16 \\
 1 8 3 & 1.12& 1.20\\
 0 7 3 & 1.13& 1.22\\
 1 8 4 & 1.14& 1.22\\
 1 9 2 & 1.18& 1.26\\
 1 9 3 & 1.20& 1.27\\
 \hline
 \end{tabular} 
\end{minipage}
\begin{minipage}{0.73\linewidth}
\scriptsize
 \begin{tabular}{|c|c|c|c|c|}

 \hline
  \multicolumn{5}{|c|}{Genetic Operations}\\
 \hline
  \multicolumn{3}{|c|}{Crossover} & \multicolumn{2}{|c|}{Mutation}\\
  \hline
  Parent 1& Parent 2& Child & Parent 1 & Child \\
 8 1 1 & 1 2 5 & (u=0.31) 8 2 5 & 8 2 5 & (u=0.34) 6 2 1\\
 8 1 1 & 1 2 5 & (u=0.81) 1 1 1 & 1 1 1 & (u=0.68) 1 1 7\\
 8 1 1 & 0 1 5 & (u=0.81) 0 1 1 & 1 1 1 & (u=0.18) 8 1 1\\
 9 1 1 & 1 2 5 & (u=0.31) 9 1 5 & 8 1 1 & (u=0.31) 2 1 4\\
 9 1 1 & 1 2 5 & (u=0.31) 9 2 5 & 8 1 1 & (u=0.15) 5 7 1\\
 9 1 1 & 0 1 5 & (u=0.31) 9 1 5 & 9 1 5 & (u=0.46) 6 1 5\\
 9 1 1 & 0 1 5 & (u=0.81) 0 1 1 & 9 1 5 & (u=0.62) 1 8 5\\
       &       &                & 9 2 5 & (u=0.81) 1 2 3\\
       &       &                & 9 2 5 & (u=0.15) 7 6 5\\
       &       &                & 9 1 1 & (u=0.50) 4 0 1 \\
       &       &                & 9 1 1 & (u=0.37) 9 2 6\\
 \hline
 \end{tabular}

\end{minipage}

Next, it is necessary to select which offers to send to the opponent. Since $p_{pevo}=100\%$, if possible, all of the offers will come from the iso-utility curve calculated using $P_{evo}$. If it is not possible, it will take as many offers as possible from the iso-utility curve from $P_{evo}$ and take the rest from the iso-utility curve from $P$.  In this case, X=(1 2 5) and Y=(0 1 5) are selected from $P$ since $P_{evo}$ does not contain elements to form a current iso-utility curve. The opponent receives the offers X and Y. Since they yield a utility of 0.25 and 0.15 respectively, both are rejected. The seller sends W=(6 1 1) and Z=(9 4 1) as counteroffers. Both of them are rejected since their utilities (0.34 and 0.18 respectively) are lower than $U_{s}(3)=0.75$.

\paragraph{Round 3 $U_{s}(2)=0.85-0.75$ $U_{b}(2)=0.85-0.75$} 
Two offers have been received from the opponent. Thus, the evolutionary sampling phase is carried out. The iso-utility curve from P ($U_{s}(2)=0.85-0.75$) and genetic operations are shown in the following tables.

\hspace{-1.5cm}\begin{minipage}{0.27\linewidth}
 \scriptsize
 \begin{tabular}{|l | l | l|}
\hline
 \multicolumn{3}{|c|}{Iso-utility curve (P)} \\
 \hline
 Offer & d(W) & d(Z) \\
 \hline
 1 3 0 & 0.60 & 0.90 \\
 1 5 0 & 0.72 & 0.90 \\
 1 5 3 & 0.74 & 0.92 \\
 \hline
 \end{tabular} 
\end{minipage}
\begin{minipage}{0.73\linewidth}
\scriptsize
 \begin{tabular}{|c|c|c|c|c|}

 \hline
  \multicolumn{5}{|c|}{Genetic Operations}\\
 \hline
  \multicolumn{3}{|c|}{Crossover} & \multicolumn{2}{|c|}{Mutation}\\
  \hline
  Parent 1& Parent 2& Child & Parent 1 & Child \\
 6 1 1 & 1 3 0 & (u=0.81) 1 3 1 & 1 3 1 & (u=0.00) 8 8 1\\
 6 1 1 & 1 3 0 & (u=0.34) 6 1 0 & 1 3 1 & (u=0.62) 1 6 7 \\
 6 1 1 & 1 5 0 & (u=0.81) 1 1 0 & 6 1 1 & (u=0.34) 6 2 1\\
 9 4 1 & 1 3 0 & (u=0.18) 9 4 0 & 6 1 1 & (u=0.21) 6 1 8\\
 9 4 1 & 1 5 0 & (u=0.81) 1 5 1 & 1 1 0 & (u=1.00) 0 1 6\\
       &       &                & 1 8 5 & (u=0.62) 1 8 5\\
       &       &                & 9 4 0 & (u=0.18) 9 5 0\\
       &       &                & 9 4 0 & (u=0.81) 1 4 1\\
       &       &                & 9 4 1 & (u=0.18) 8 4 1\\
       &       &                & 9 4 1 & (u=0.18) 9 6 1\\
       &       &                & 1 5 1 & (u=0.62) 1 7 0\\
       &       &                & 1 5 1 & (u=0.31) 4 7 1\\
 \hline
 \end{tabular}

\end{minipage}

Next, it is necessary to select which offers to send to the opponent. The table below shows the iso-utility curve calculated from $P_{evo}$.  In this case, X=(1 1 1) and Y=(1 1 0) are selected from $P_{evo}$. The opponent receives the offers X and Y. Since they yield a utility of 0.69 and 0.53 respectively, both are rejected. However, in this round, the seller sends W=(4 1 1) as counteroffer. The offer is rejected because its utility is equal to 0.5, and is thus lower than  $U_{b}(4)=0.65$. From this point on we will overlook the inner steps of the model due to the fact that the way it works has already been described.
\begin{center}
\begin{tabular}{|l | l | l|}
\hline
 \multicolumn{3}{|c|}{Iso-utility curve ($P_{evo}$)} \\
 \hline
 Offer & d(W) & d(Z) \\
 \hline
 1 1 0 & 0.56 & 0.95 \\
 0 1 1 & 0.66 & 1.05 \\
 1 4 1 & 0.64 & 0.88 \\
 1 3 1& 0.59 & 0.89\\
 1 5 1& 0.71 & 0.89\\
 1 1 1& 0.55 & 0.94 \\
 1 2 3 & 0.60 & 0.94\\
 \hline
 \end{tabular}
\end{center}

\paragraph{Round 4 $U_{s}(2)=0.75-0.65$ $U_{b}(2)=0.75-0.65$} In this round, the buyer sends X=(1 1 7), which yields a utility of 0.33 for the seller. Therefore, the offer is rejected. Then, the opponent sends W=(1 1 1) and Z=(1 2 1), Z being accepted by the buyer since its utility is equal to 0.81. The negotiation process ends with the deal ($U_{b}=0.81$,$U_{s}=0.69$), which is the Nash Bargaining Point for this negotiation case. 

This section has described the main traits of the proposed negotiation model for AmI environments. More specifically, it has explained the protocol employed, and the negotiation strategy that is adapted to AmI domains thanks to the intelligent sampling provided by genetic operators during the negotiation process. Additionally, we have also shown how the proposed model works in a small case study. In the next section the proposed model is tested in several scenarios to check its performance.

\section{Experiments}
\label{exp}
The performance of the devised strategy is detailed in this section. The proposed negotiation model was tested against the weighted constraint model proposed by Ito et al. \cite{ito08}. This model makes it possible to represent unrestricted interdependence relationships among the negotiation issues. Furthermore, if the number of constraints is large, it can represent highly non-linear utility functions. Therefore, it represents a proper testbed for the proposed strategy. Nevertheless, as in the work of Lai et al. \cite{lai08}, the proposed negotiation model is general and does not depend on a particular utility function. The model of Ito et al. was selected as a testbed because it provides a well studied utility function \cite{ito08,marsa09,marsa09b} that holds enough complexity to study the real performance of the negotiation model. 

Firstly, the negotiation setting employed in the experiments is briefly described. After this, the different experiments and their results are presented. Finally, a brief discussion summarizing the results of the experiments is included.   
\subsection{Negotiation setting}

The aim of these experiments was to evaluate whether or not the proposed model is capable of working in domains where the agents' utility functions are highly non-linear. For that purpose, different negotiation cases where randomly created:
\begin{itemize}
 \item Number of issues $N$ = [4-7].
 \item Integer issues. $s_{i} \in [0,9]$.
 \item $L=N*$5 uniformly distributed constraints per agent. There are constraints for every possible interdependence cardinality. For instance if $N$=4, there are 5 unary constraints, 5 binary constraints, 5 trinary constraints and 5 quaternary constraints.
 \item $v(c_{l},.)$ for each $n$-ary constraint  drawn randomly from $[0,100*n]$. 
 \item For every constraint, the constraint width for each issue $s_{i}$ is uniformly drawn from $[2,4]$. For instance, if the constraint width for issue $s_{1}$ is 3, then $(0\leq s_{1} \leq 3)$, $(1\leq s_{1} \leq 4)$, $(2\leq s_{1} \leq 5)$, $(3\leq s_{1} \leq 6)$, $(4\leq s_{1} \leq 7)$, $(5\leq s_{1} \leq 8)$ and $(6\leq s_{1} \leq 9)$ are all of the possible configurations for issue $s_{1}$ in the constraint (just one is used in the constraint).
 \item Agent deadline $T=10$. Agents do not know their opponent's private deadlines.
 \item Agent reservation utility $RU=0$. Agents do not know their opponent's private reservation utilities.It is set to zero in order to find a deal, if possible. Should this be the case, the deal is checked against certain thresholds which will determine whether the application notifies its owner of the possible deal.
 \item Agents do not know their opponent's utility functions
\end{itemize}

For each number of issues, a total of 100 negotiation cases were generated with the above settings. The execution of each case was repeated 30 times in order to allow for the possible differences between different executions of the methods. 

In order to evaluate the quality of the agreements found by the participant agents, some measures were gathered at the end of each negotiation.
\begin{itemize}
 \item Euclidean distance to the closest Pareto frontier point \cite{osborne99}. This is a measure of economic efficiency for agreements. The closer to the Pareto frontier, the better.
 \item Euclidean distance to the Nash Product \cite{osborne99}. Since both agents have the same concession strategy and the same deadline it is also feasible to study the distance to the Nash Product. This is the point that maximizes the product $u_{1}*u_{2}$ in the Pareto Frontier, where $u_{1}$ is the utility of agent 1, and $u_{2}$ is the utility of agent 2.
 \item Number of negotiation rounds. Faster agreements are preferred since a lesser number of messages are exchanged, less bandwidth is needed, and limited devices need less power to send messages.
\end{itemize}

Additionally, some experiments were also devised in order to test the computational performance of the proposed model in a real environment. Measures such as the time spent in decision making tasks before the negotiation process (self-sampling) and during the negotiation process (opponent offer acceptance phase, evolutionary sampling, and offer proposal) were gathered. For that purpose, the proposed model was implemented using a HTC Desire (1 Ghz, 576MB RAM, Android Operating System) as one of the parties and a PC (2 Ghz, 4096MB RAM, Ubuntu Operating System) as the other party. A total number of 30 negotiations were carried out in order to measure the computational cost of the proposed model.

\subsection{Results}

 The proposed strategy, which will be named as \textit{Evolutionary Sampling} or \textit{ES}, was compared with two different negotiation models. The first strategy is an implementation of the general framework proposed by Lai et al. \cite{lai08}. This model is provided with the whole sampling of the utility function, so that it can completely calculate iso-utility curves. It is used as a measure of how close the proposed strategy is to the ideal case where all of the offers are available. The second model assumes that it is not possible to completely sample all of the offers. Therefore, it samples before the negotiation process by means of a niching GA (\textit{self-sampling}) and uses the similarity heuristic ($p_{pevo}=0$) during the negotiation process, which will be named as \textit{Non Evolutionary Sampling} or \textit{NES} model. The number of samples explored by the \textit{NES} model before the negotiation process is set equal to the number of samples explored by the ES model ($|P|+Samples_{evo}$). Consequently, both the \textit{NES} and \textit{ES} model yield the same computational cost in every experimentation.

 Four different experiments were carried out in order to test the proposed model. In the first experiment, the three different models are compared as the number of issues is increased. The second experiment, studies the impact of the proportion of offers ($p_{pevo}$) that are sent from the special pool $P_{evo}$ in the \textit{ES} model. Next, the three models are compared as the number of proposals $k$ increases. Finally, the \textit{ES} and the \textit{NES} model are compared as the size of the population ($|P|$) provided by the \textit{self-sampling} increases.

\subsubsection{Experiment 1: Performance study on the number of issues}

The goal of this experiment is to study how the proposed strategy behaves for negotiations with a different number of issues $N=\{4,5,6,7\}$. It is important that the proposed model be capable of properly handle negotiations with multiple issues since most real world domains, including AmI domains, need to reach agreements for multiple issues. A negotiation setting where agents are limited to $k=3$ proposals per negotiation round is used. The three different models were tested during this experiment.

The parameters of the \textit{self-sampling} were set to $n_{max}=100$, $p_{dc}=80\%$ and $p_{cr}=80\%$. The number of samples optimized before the negotiation process was set to $|P|=128$ for the \textit{ES} model and to $|P|=128+Samples_{evo}$ for the \textit{NES} model.

The parameters of the \textit{ES} were set to $M=5$, $n_{cross}=4$, $n_{mut}=4$, $p_{attr}=30\%$, and $p_{pevo}=100\%$. Therefore, all the offers are sent from the samples generated by the \textit{evolutionary sampling} carried out during the negotiation process. 

\begin{figure}[ht]

\begin{tabular}{c c}
 \includegraphics[scale=0.65]{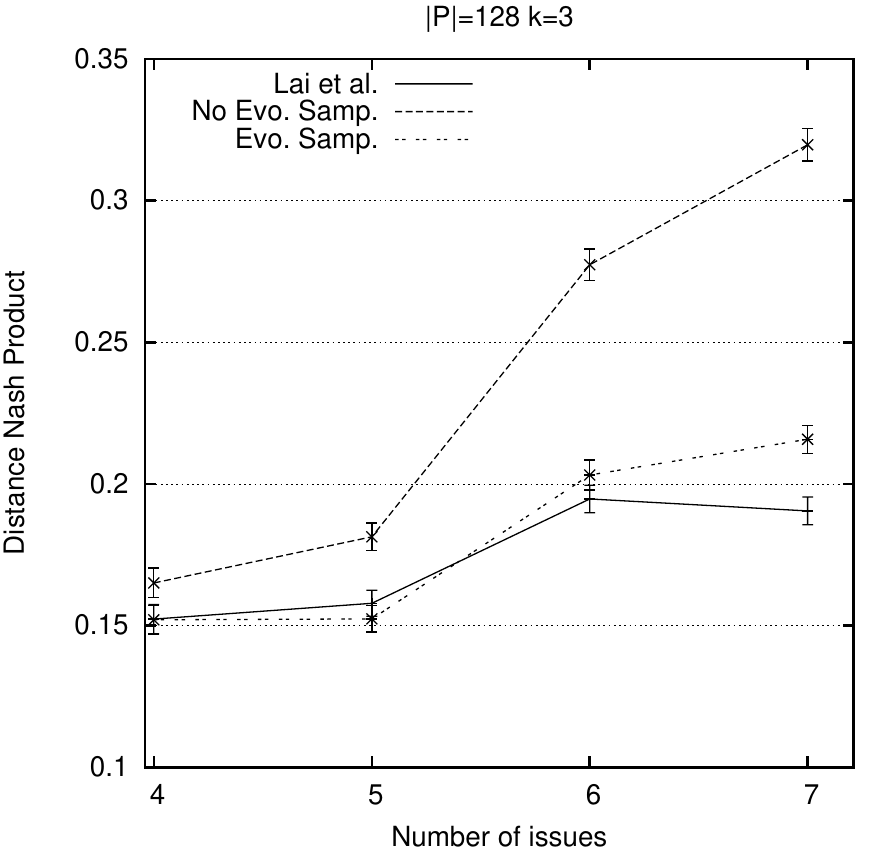} & \includegraphics[scale=0.65]{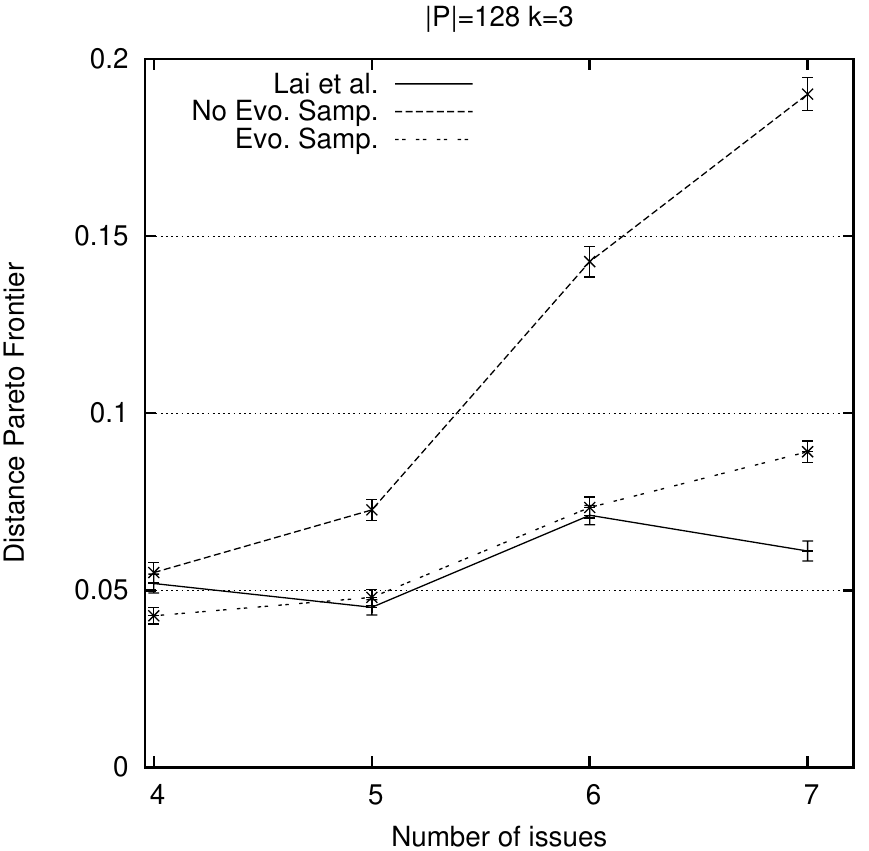}
\end{tabular}
 \centering
 \includegraphics[scale=0.65]{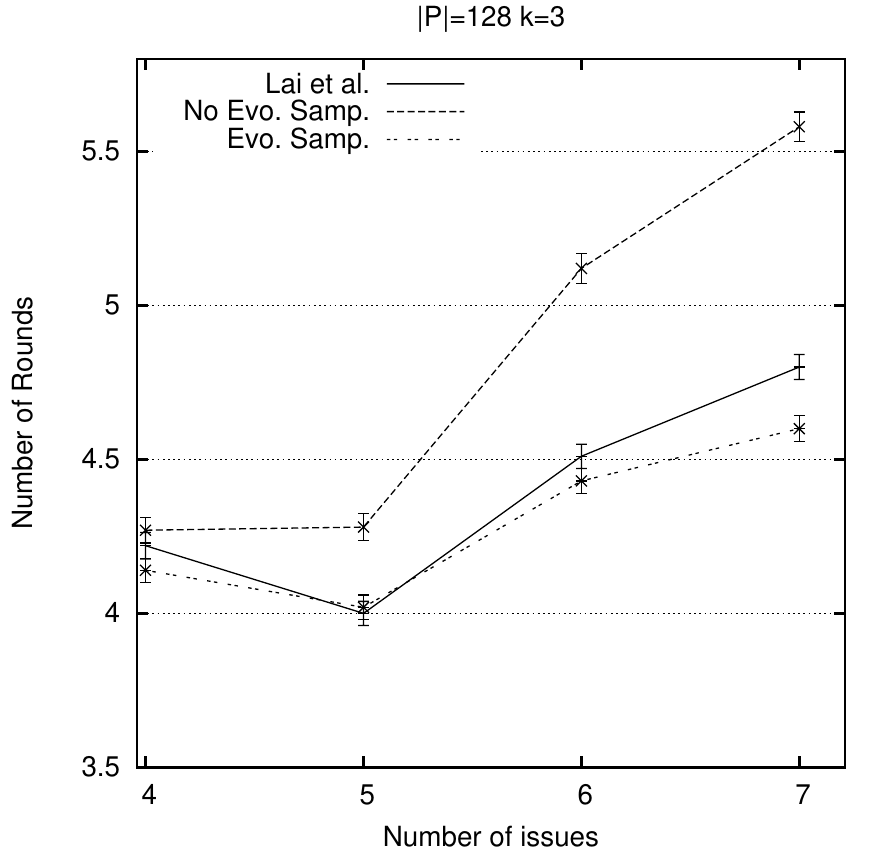}
 \caption{Evolution of the distance to the Nash Product, distance to the closest Pareto Point, and number of negotiation rounds in Experiment 1. The graphic shows the mean and its associated confidence intervals (95\%) }
\label{fig:exp1}
\end{figure}

The distance to the Nash Product, the distance to the closer Pareto Frontier Point and the number of negotiation rounds were measured for the three models. The results for this experiment can be found in Figure \ref{fig:exp1}. Intuitively speaking,  since the number of offers sampled remains constant and the number of issues increases, the performance of the \textit{NES} and the \textit{ES} model should be worsened with respect to the results achieved by the model of Lai et al. However, the results for the \textit{ES} do not comply with this intuitive hypothesis.  As can be observed, even though the proposed model and the \textit{NES} model explore the same number of offers, the \textit{NES} obtains worse results than the other two models. This is particularly true as the number of issues increases, since the performance of this method drastically decreases. On the contrary, the \textit{ES} model is capable of achieving statistically equal results to the model of Lai et al., which can access the whole iso-utility curve. Nevertheless, the proposed model explores far fewer offers than the complete sampling of the utility function, especially for larger number of issues. For instance, when $N=6$, Lai et al. has access to $10^{6}$ offers, whereas the proposed model has only sampled an average of 1510 samples (128+ average $Samples_{evo}$). 

The \textit{ES} model has been able to achieve similar results to the case where the full iso-utility curve can be calculated, while maintaining the offers sampled to a small number. This result is particularly interesting for AmI domains where agents may be executed in devices with low computational and storage capabilities. Therefore, fewer samples mean less power consumption and less capacity needed to store them. Moreover, it must also be highlighted that the number of rounds was also lower than that obtained by \textit{NES}, which,  consequently means fewer messages sent, less bandwidth needed and, of course, less power consumption by the limited devices.

The reason for this improvement is the intelligent sampling achieved by the use of genetic operators during the negotiation process. On the contrary, sampling only before the negotiation process leads to worse results since it is not capable detecting which offers will be interesting for the negotiation. Both, the \textit{ES} and the \textit{NES} model, have the same computational cost, but the \textit{ES} is obviously preferred since it is capable of achieving a better performance.

\subsubsection{Experiment 2: Performance study on $p_{pevo}$}

In this case, the experiment's goal is to study how relevant the proportion of offers that are sent from the offers sampled during the negotiation process (governed by the parameter $p_{pevo}$) in the \textit{ES} model is. Since all of the configurations sample new offers during the negotiation process, all of them yield a very similar computational cost. In fact, it may only be different if one of the configurations obtains a significantly different number of negotiation rounds. Consequently, the main subject of study in this scenario is the economic efficiency (distance to Nash and Pareto Frontier), although some improvements in the computational cost may be observed due to a lower number of rounds.

The same conditions from the previous experiment were set ($k=3$ and $N=\{4,5,6,7\}$), and the same configuration parameters were set for the \textit{ES} ($M=5$, $n_{cross}=4$, $n_{mut}=4$, and $p_{attr}=30\%$). However, in this scenario we compare the \textit{ES} model results when 1 out of 3 offers ($p_{pevo}=30\%$), 2 out of 3 offers ($p_{pevo}=50\%$), and 3 out of 3 offers ($p_{pevo}=100\%$) come from the offers sampled during the \textit{evolutionary sampling} phase. 

\begin{figure}[ht]
\begin{tabular}{c c}
 \includegraphics[scale=0.65]{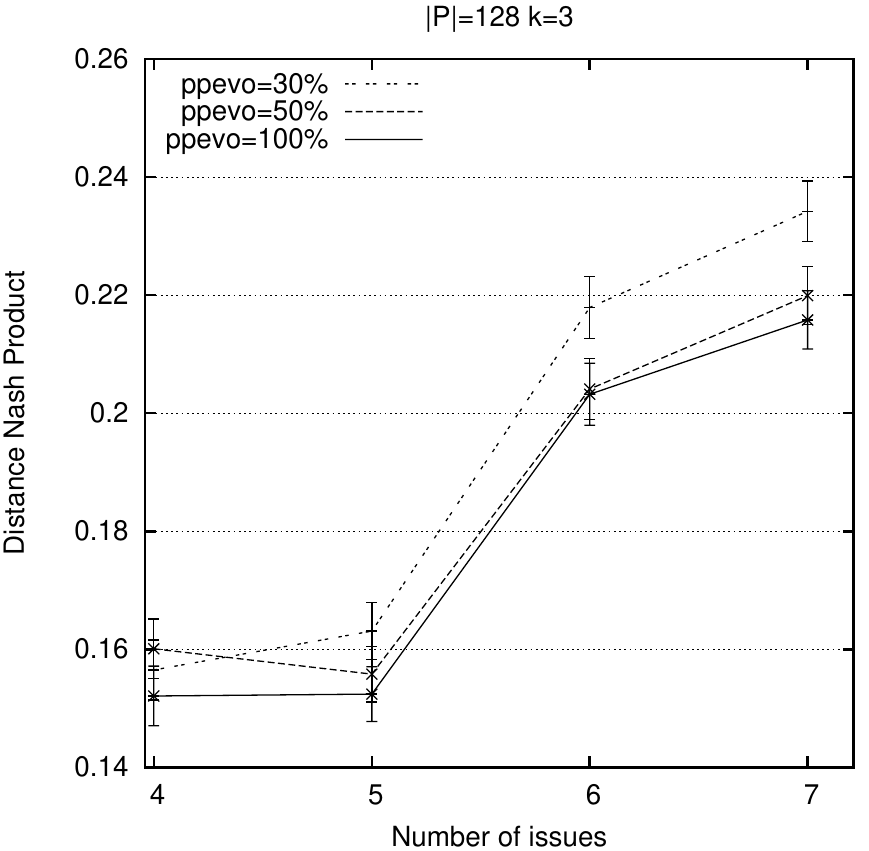} & \includegraphics[scale=0.65]{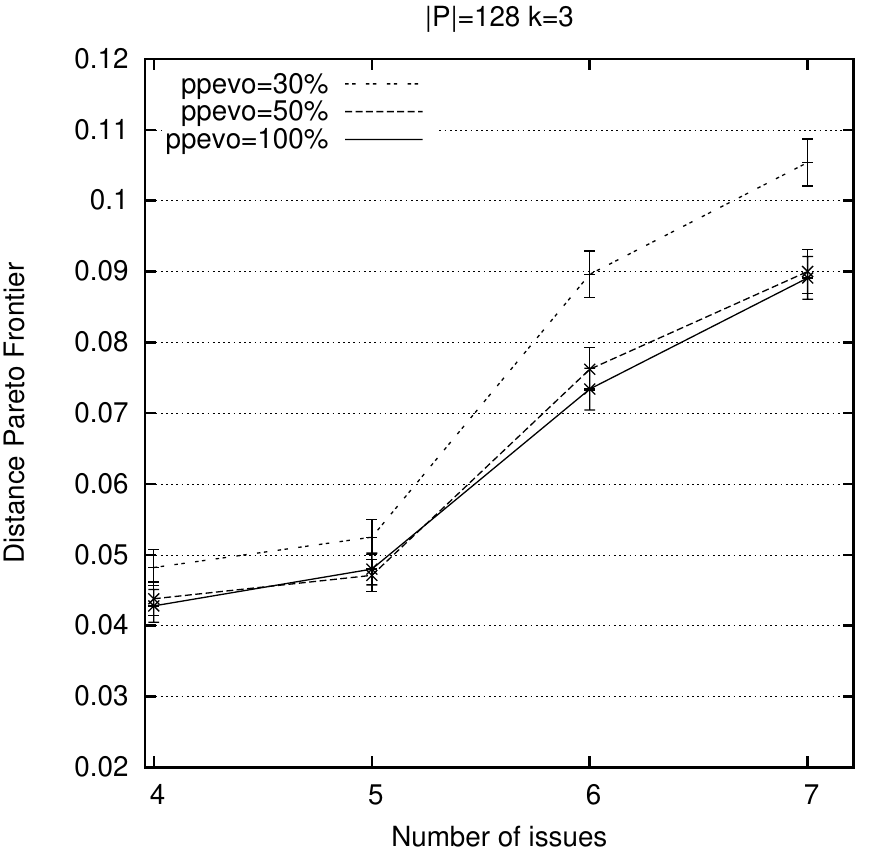}
\end{tabular}
 \centering
 \includegraphics[scale=0.65]{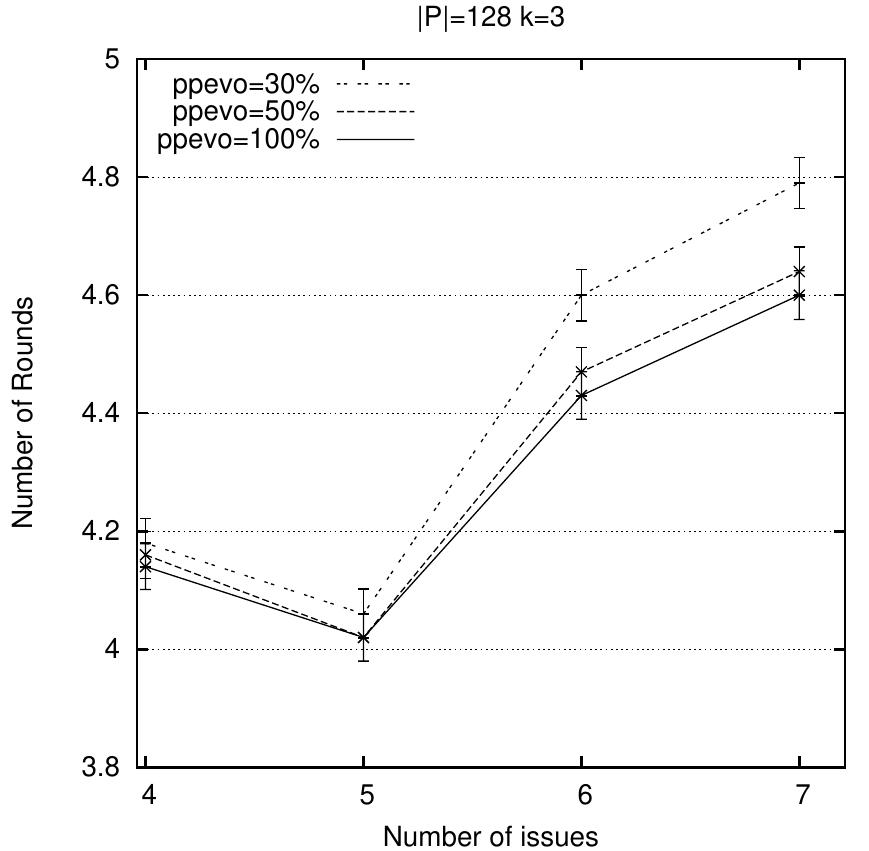}
 \caption{Evolution of the distance to the Nash Product, distance to the closer Pareto Point, and number of negotiation rounds in Experiment 2. The graphic shows the mean and its associated confidence intervals (95\%) }
\label{fig:exp2}
\end{figure}

The results for this second scenario can be observed in Figure \ref{fig:exp2}. The graphic shows that the three different configurations yield similar results for the distance to the Nash Product, the distance to the closest Pareto Frontier Point, and the number of negotiation rounds. This similarity is explained due to the fact that, on most occasions, the offer accepted by the opponent is the closest one from the \textit{evolutionary sampling} population ($P_{evo}$). Therefore, it is always sent, as long as the results from the \textit{evolutionary sampling} are not ignored. Nevertheless, it seems that higher values of $p_{pevo}$ have a slightly better economic and computational performance than lower ones. The reason for this slight improvement is that,  in some cases, the offer preferred by the opponent may be the second or third closest from $P_{evo}$. Due to this small improvement, higher values of $p_{pevo}$ are preferred in practice.  

\subsubsection{Experiment 3: Performance study on $k$}

The next experiment aims to study the performance of the three different models (Lai et al., \textit{NES}, and \textit{ES}) as the limit to the number of offers $k$ sent per agent's round is increased. As mentioned, the number of offers sent may help to reach agreements faster since agents are capable of finding \textit{win-win} situations. This is very important in AmI environments where devices have limited power and their running time must be optimized. Lai et al. \cite{lai08}, demonstrated how higher values of $k$ helped to reach better agreements. In this scenario, the experiment is repeated in order to evaluate whether the differences between the three models still hold for different values of $k$. 

The studied values of $k$ were 1, 3, 5, and 7. The rest of the negotiation setting was configured to use negotiation cases with $N=6$ issues. The parameters of the \textit{self-sampling} were set to the values employed in the previous tests except for $|P|=256$. The parameters of the \textit{ES} were set to the same conditions described in Experiment 1. 

\begin{figure}[ht]

\begin{tabular}{c c}
 \includegraphics[scale=0.65]{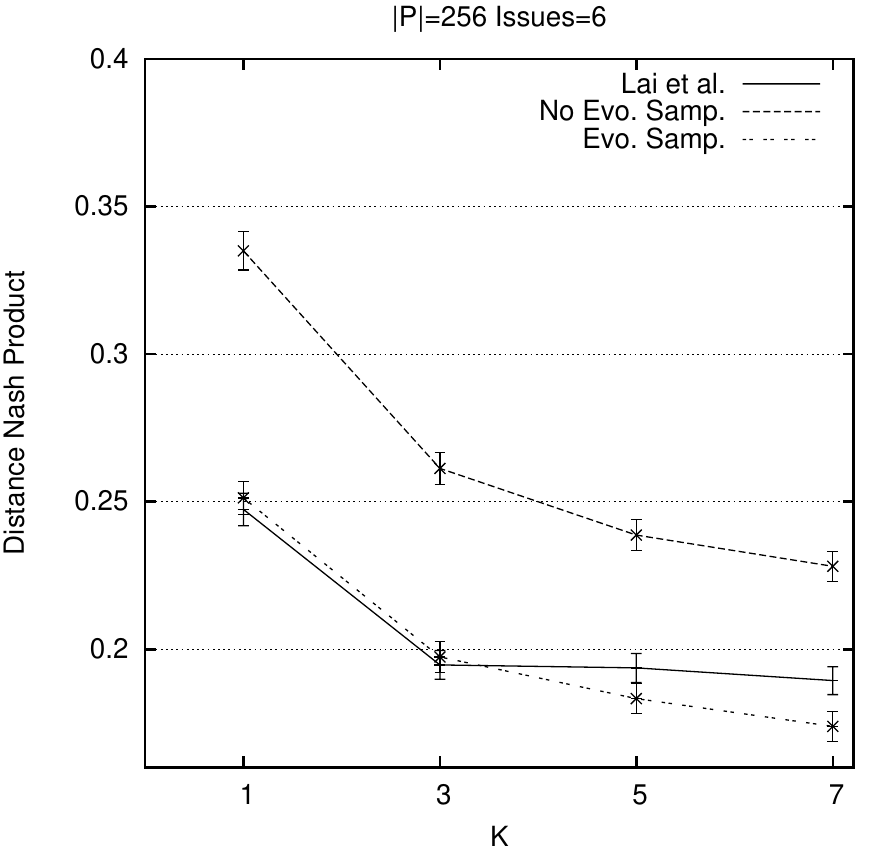} & \includegraphics[scale=0.65]{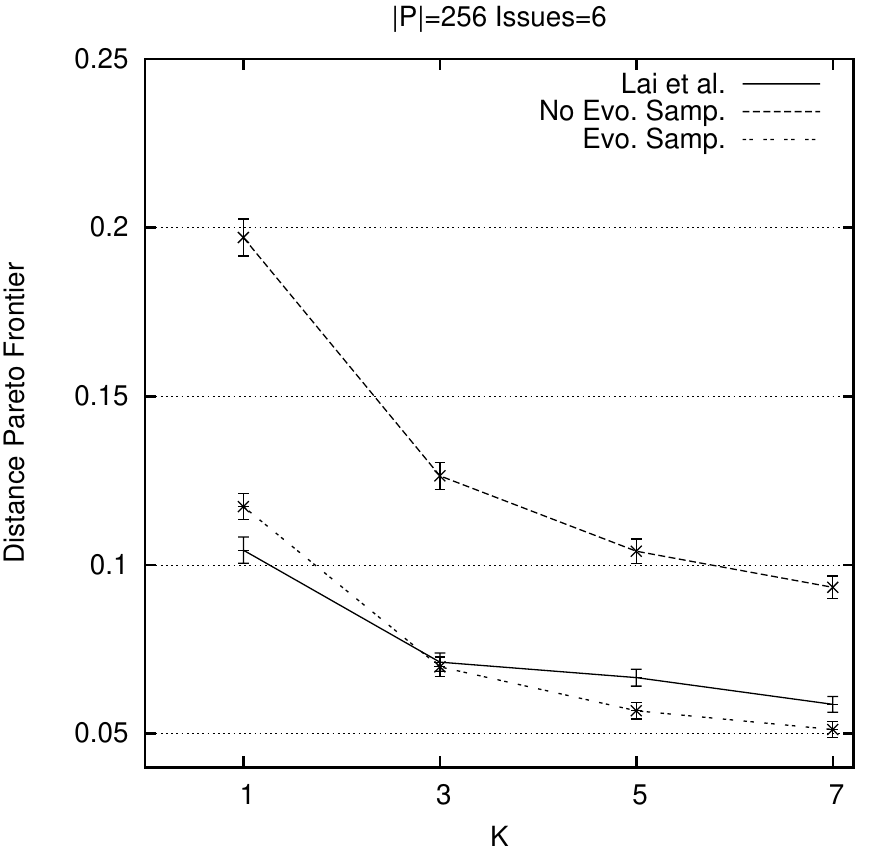}
\end{tabular}
 \centering
 \includegraphics[scale=0.65]{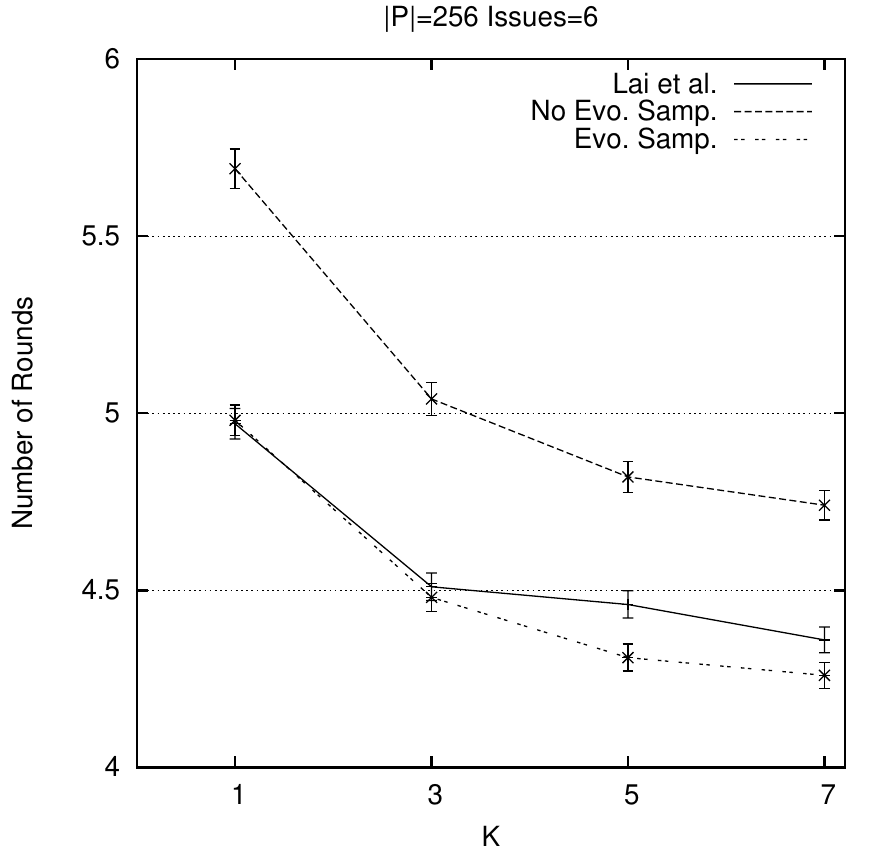}
 \caption{Evolution of the distance to the Nash Product, distance to the closest Pareto Point, and number of negotiation rounds in Experiment 3. The graphic shows the mean and its associated confidence intervals (95\%) }
  \label{fig:exp3}
\end{figure}

As it can be observed in Figure \ref{fig:exp3}, the three models achieve better results as $k$ increases. These results agree with those presented in \cite{lai08}. Although all of the models improve, the differences observed in Experiment 1 still hold for this scenario. The \textit{NES} model gets worse results than Lai et al. and the proposed model. On the contrary, the \textit{ES} obtains results that are statistically equivalent to the case when the full iso-utility curve can be calculated. As a matter of fact, for higher values of $k$ the proposed model gets slightly better results than Lai et al. Nevertheless, the differences between the two of them are not significant enough to be considered as relevant.

It must be noted again that the number of offers sampled for \textit{ES} and \textit{NES} is the same and it is much lower than the complete sampling of the utility function. For instance, in this scenario, the complete sampling consists of $10^6$ offers, whereas the other two methods sampled an average of 773 samples for $k=1$, 1653 for $k=3$, 2497 for $k=5$, and 3357 for $k=7$.

\subsubsection{Experiment 4: Performance study on $|P|$ and memory performance}

This last experiment was designed to assess the influence of the population optimized by the \textit{self-sampling} on the performance of the \textit{ES} model and the \textit{NES} model. It is specially relevant to see how many samples the \textit{NES} model needs to achieve similar results to those ones obtained by the model proposed in this article. Obviously, more population means more storage needed and more computational cost since it needs to optimize more samples.

The average number of samples explored was analyzed for a negotiation setting where $N=6$ and $k=3$. The settings used for the \textit{self-sampling} and the \textit{ES} in previous experiments were repeated for this scenario. The number of sampled offers was increased by allowing more offers to be optimized in the \textit{self-sampling} ($|P|={128,256,512,1024,2048,4096}$). 

\begin{figure}[ht]
\hspace{-2cm}
\begin{tabular}{c c}
 \includegraphics[scale=0.65]{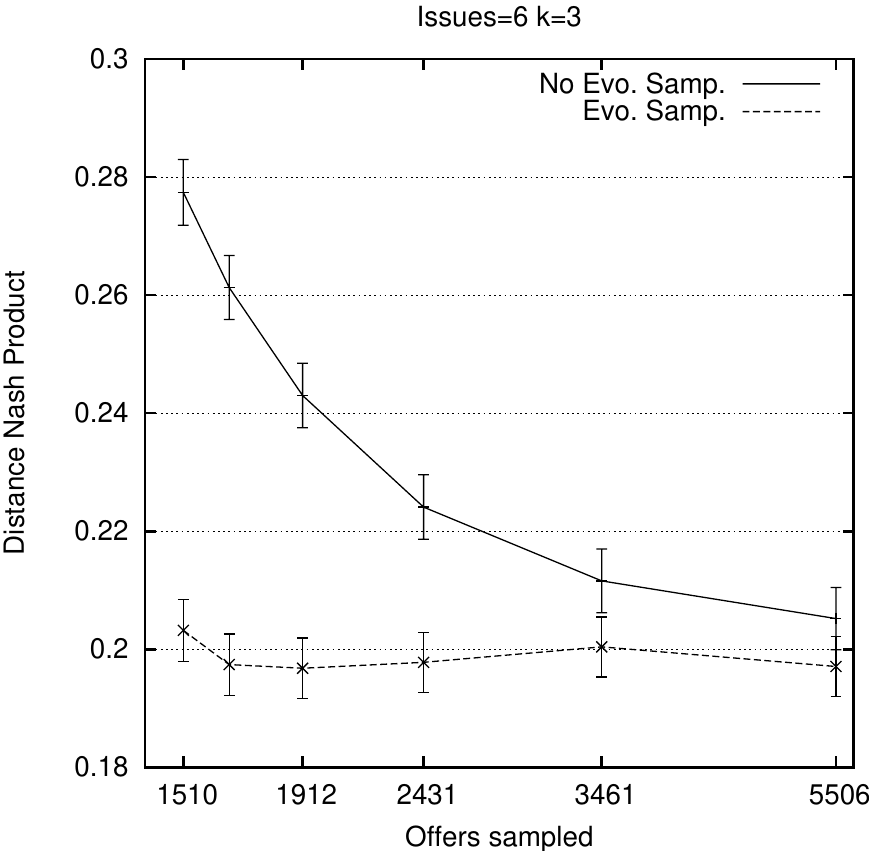} & \includegraphics[scale=0.65]{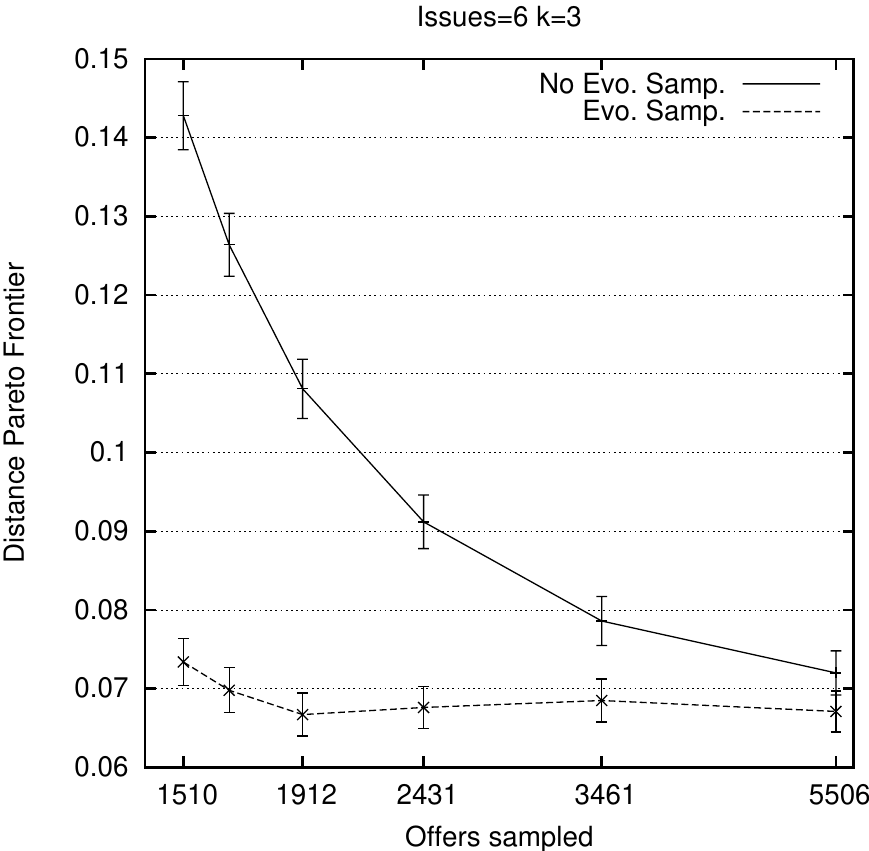}
\end{tabular}
 \centering
 \hspace{-2cm}
\includegraphics[scale=0.65]{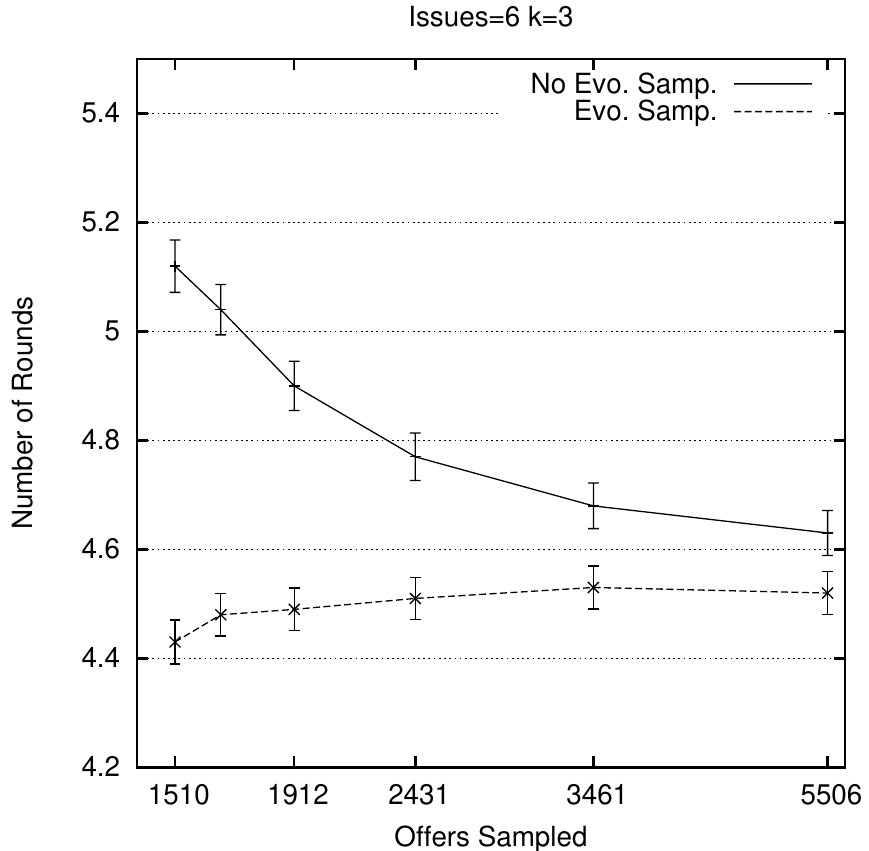}
 \caption{Evolution of the distance to the Nash Product, distance to the closest Pareto Point, and number of negotiation rounds in Experiment 4. The graphic shows the mean and its associated confidence intervals (95\%) }
 \label{fig:exp4}
\end{figure}

The results for this experiment can be observed in Figure \ref{fig:exp4}. The x axis of the graphics show the average number of offers sampled by both models, thus it shows $|P|+average_{rounds}*Samples_{evo}$. In the case of the \textit{NES} model all of the samples were produced before the negotiation process started. Several observations can be made from the data shown in the graphics. On the one hand, it seems that the size of $|P|$ does not have too much of an effect on the performance of the \textit{ES} model, since it is more dependent on the exploration carried out during the negotiation process and does not need as much sampling to get results similar to the case where the full iso-utility curve can be accessed. Therefore, the behavior of the model remained almost constant for different configurations of $|P|$. Again, this behavior is very adequate for AmI environments since the model can properly work with configurations that do not require too many computational resources.  On the other hand, the \textit{NES} model performance increased along with the number of offers sampled. It must be noted, that when the number of samples for both methods was 5506, the two of them obtained very similar, almost equivalent, results. Therefore, the \textit{NES} needed 5506 samples to achieve similar results to the same results obtained by the \textit{ES} model for 1510 samples. It can be concluded that \textit{NES} needs $\frac{5506}{1510}=3.64$ times more samples to achieve similar results to \textit{ES}.

It is possible to approximately analyze the total amount of memory employed by both methods when they achieve statistically equivalent results. As has been suggested by the previous experiment (Experiment 4), the \textit{NES} model needs 5506 samples to achieve statistically equivalent results to those the \textit{ES} model with 1510 samples. If we assume that the underlying platform is a 32 bit platform, where integers usually need 32 bits to be stored, we can approximately calculate the memory needed by both models as follows:

\begin{equation}
 Memory(KB)= |\mbox{Samples}|*N*32*\frac{1}{8}*\frac{1}{1024}
\end{equation}
where $|\mbox{Samples}|$ is the number of samples, $N$ is the number of attributes of the negotiation process, 32 is the size of an integer, $\frac{1}{8}$ converts from bits to Bytes, and $\frac{1}{1024}$ converts from bytes to KBytes. Taking into account the formula above, the \textit{NES} model would take 129 KB to store the data needed for the previous type of negotiation process ($N=6$, $|Samples|=5506$), whereas the \textit{ES} model would take 35 KB ($N=6$, $|Samples|=5506$). Depending on the underlying device, this difference may be quite important (e.g. devices with a few MB of storage available). However, this difference may be still more important if we consider that in some scenarios it may be necessary to perform several negotiations at the same time (e.g. the fair scenario). For instance, Table \ref{comp:mem} shows the approximate amount of memory necessary to carry out several negotiations at the same time.

\begin{table}
\begin{center}
 \begin{tabular}{|l|l|l|}
\hline
Number negotiations & Memory(KB) \textit{NES} & Memory(KB) \textit{ES} \\
\hline
1 & 129 & 35 \\
\hline
3 & 387 & 105 \\
\hline
5 & 645 & 175 \\
\hline
7 & 903 & 245 \\
\hline
10 & 1290 & 350 \\
\hline
 \end{tabular} 
 \end{center}
\caption{Approximate ammount of memory needed by the \textit{NES} model and the \textit{ES} model when executing several negotiations at the same time}
\label{comp:mem}
\end{table} 

\subsubsection{Experiment 5: Time Performance}

As introduced earlier, it was also interesting to test the computational performance of the model in a real environment. Thus, the proposed model was implemented using a HTC (1 Ghz, 576MB RAM, Android Operating System) as one of the parties and a PC (2 Ghz, 4096MB RAM, Ubuntu Operating System) as the other party. The \textit{self-sampling} parameters were set to $n_{max}=100$, $p_{dc}=80\%$ and $p_{cr}=80\%$. The number of samples optimized before the negotiation process was set to $|P|=128$. As for the parameters employed during the negotiation process, these were set to $k=3$, $M=5$, $n_{cross}=4$, $n_{mut}=4$, $p_{attr}=30\%$, and $p_{pevo}=100\%$. The number of attributes of the negotiation process was $N=5$. The time spent in the whole negotiation process ($t_{t}$), the time spent in sending/waiting for offers ($t_{m}$), the time spent in \textit{self-sampling} ($t_{s}$), and the time spent in decision-making during the negotiation process ($t_{dm}$) were measured. Table \ref{tab:comp} shows the mean obtained in seconds for the 30 negotiation cases that were studied.

\begin{table}
\begin{center}
  \begin{tabular}{|l|l|l|l|}
\hline
  $t_{t}$ (s) & $t_{s}$ (s) & $t_{dm}$ (s) & $t_{m}$ (s) \\
\hline
 0.773 & 0.264 & 0.358 & 0.415 \\ \hline
 \end{tabular} 
\end{center}
\caption{Computational performance measures for Experiment 5}
\label{tab:comp}
\end{table} 

As can be observed, the time spent for a negotiation process $t_{t}$ was reasonably good (less than a second) and it enables negotiations to be carried out in environments where real-time responses are needed (e.g. Ambient Intelligence). Moreover, it can also be observed that the time spent in decision-making tasks $t_{dm}$ does not take as much time as other tasks such as sending/waiting for offers $t_{m}$. This leaves room for more negotiation processes to be carried out in parallel during CPU idle time (e.g. waiting for offers). Again, carrying out multiple negotiation processes simultaneously proves especially interesting again for AmI environments. For instance, in the fair scenario, it makes it possible to negotiate simultaneously with those vendors who are available in the area where the consumer is walking at that moment.  The time spent in self-sampling is the least problematic since it is a process to be carried out only once until agent preferences change. In some AmI environments, such as the fair, we may consider preferences to be static during the fair event. Thus, \textit{self-sampling} would only be needed once. Despite all those facts, it must be remarked that the time spent in \textit{self-sampling} is reasonably good (less than a second).

\subsection{Discussion}

Ambient Intelligence domains are characterized as domains where computational resources are of extreme importance. Users interact with its environment through devices with limited capabilities, thus the efficient use of resources is crucial. Furthermore, the environment infrastructures are usually connected by means of a limited bandwidth wireless connection. Thus, network resources must also be optimized. 

The results obtained by the proposed model, while maintaining fairly good economic performance, cope with the problems found in AmI environments. If we assume that limited devices cannot completely sample the agent's utility function and store those samples, some mechanisms are needed to sample as few offers as possible. A straightforward method would be sampling some offers before the negotiation process starts, which is precisely what the \textit{NES} model does. However, this sampling does not take advantage of the information revealed by the opponent in the negotiation process. Most of the offers sampled before the negotiation process may be useless since they are of no interest to the opponent. However, the proposed model takes advantage of this information and employs it to make a more intelligent sample, optimizing the computational resources. Nevertheless, although computational resources are important, economic efficiency should not be ignored in AmI negotiations. 

In the previous sections, we were able to observe the behavior of the \textit{ES} model in different scenarios. Its performance was compared with a method that samples the same number of offers before the negotiation process (\textit{NES}), and the ideal case where all of the samples of the utility function are available. The results of the experiments can be summarized as:

\begin{itemize}
 \item The proposed model needs very few computational resources and storage to obtain results statistically equivalent to the ideal case where the all of the offers are available.  It obtained similar results in economic performance (distance to Nash, distance to Pareto Frontier) and number of negotiation rounds. 
 \item When the proposed model and the \textit{NES} model sample the same number of offers, the first obtains better results. In fact, the \textit{NES} model needs to sample 3.64 times more offers to obtain similar results.
 \item The proposed model needs less negotiation rounds to achieve better results than the \textit{NES} model. Therefore, the environment bandwidth is optimized since it needs fewer messages to be sent in order to reach agreements.
\end{itemize}

Consequently, the proposed model fits perfectly for the conditions needed by AmI environments, since it needs less computational resources and it obtains economically efficient results.

\section{Related Work}
\label{related}

Ambient Intelligence looks to offer personalized services and provide users with easier and more efficient ways to communicate and interact with other people and systems \cite{corchado08,weber05}. Since several users may coexist in AmI environments, it is quite probable that their preferences conflict and thus mechanisms are needed to allow users to cooperate. For instance, imagine a ubiquitous shopping mall \cite{keegan08,bajo09} where buying agents have to help users to buy the products, and vendor agents have to maximize their users' profits. Automated negotiation provides mechanisms that solve this particularly interesting problem. Some authors have already claimed that in most real world negotiations such as e-commerce \cite{klein02,robu05,ito08}, issues present interdependence relationships that make agents' utility functions complex. Therefore, the problem of complex utility functions in automated negotiation is also interesting for AmI applications.

Over the last few years, most of the work in automated negotiation has focused on offering solutions for the case of imperfect knowledge and bounded computational resources \cite{kraus97,jennings01}. The use of \textit{heuristics} is necessary to provide a solution to problems of this type. The present work can be classified within this same category of solutions.  

Faratin et al. \cite{faratin98} presented a negotiation model for linear utility functions where a negotiation strategy is composed of different tactics that may be applied depending on the negotiation time, the quantity of the resource and the behavior of the opponent. Nevertheless, the model is only applicable in negotiation with linear utility functions, which are easier cases than those presented in this present article. 

Matos et al. \cite{matos98} determined the successful strategies for different settings using the model proposed by Faratin et al. \cite{faratin98}. They employ an evolutionary approach in which strategies and tactics correspond to the genetic material in a genetic algorithm. In their experiments, populations of buyers and sellers with different strategies negotiate in a round robin way. After each round robin round, strategies are evaluated by means of a fitness function. Then, strategies are selected to be the parents of the next population according to their fitness function. In the end, a population of strategies implicitly adapted to the environment is obtained. They use genetic algorithms as a learning mechanism of negotiation strategies when placed under certain circumstances. There are two differences between Matos et al. work and the present work. Firstly, the negotiation model of Matos et al. is designed for linear utility functions. Secondly, the genetic algorithm proposed in this present work is an implicit learning mechanism of the opponent's preferences that guides the offer sampling during the negotiation process.  

Later, Faratin et al. \cite{faratin00} presented a negotiation strategy for bilateral bargaining that is focused on achieving \textit{win-win} situations by means of trade-off. The heuristic applied to perform trade-off is similar to that employed in this present work. Given an agent's current utility, the offer from the iso-utility curve that is most similar to the last offer received from the opponent is sent. The idea behind this heuristic is that, since the proposed offer is the most similar to the last offer received from the opponent, it is more likely to be satisfactory to both participants. A fuzzy similarity criterion is employed to compare offers. Nevertheless, the use of fuzzy similarity requires some knowledge of opponent preferences. The application of criteria of this kind is complicated in complex utility functions due to the interdependencies among the different issues. In this present work, the Euclidean distance is used, as this does not require any knowledge about the opponent, and which is independent of the interdependencies among issues.

Fatima et al. \cite{fatima03,fatima04a,fatima04b} analyzed the problem of multi-attribute negotiations in an agenda-based framework. Agendas determine in which order the different issues are to be negotiated when negotiations are carried out issue by issue. Once an agreement has been found on a specific issue, it cannot be changed. Thus, the agents face the problem of which issues should be negotiated first and which strategies should be applied. They studied the optimal agendas for different scenarios. Nevertheless, their work focused on linear utility functions, which does not take into account the possible interdependences among the different issues.

 The work of Krovi et al. \cite{krovi99} opened the path for GA's in automated negotiation. Krovi et al. proposed a GA for bilateral negotiations that was performed each time a negotiation round ended. The population of chromosomes was randomly initialized with 90 random offers and 10 heuristic offers (the last offer from the opponent and the nine best offers from the previous round). The idea behind using GA's is that the resulting offers have good characteristics for both agents. However, 60 generations were needed during each round in order to obtain the next offer, which may turn out to be computationally expensive in large issue domains. Choi et al. \cite{choi01} enhanced Krovi's model with more learning capabilities. More specifically, it is capable of learning opponent preferences by means of stochastic approximation and of adapting its mutation rate to opponent behavior. However, these strategies and mechanisms are devised for linear utility functions with few negotiation issues. The performance of these methods is uncertain when a large number of issues or complex utility functions are used. This present work also employs genetic operators to obtain new offers, but it is capable of providing solutions for domains with complex utility functions and domains where the number of issues is large.

 There have been some works that have studied the problem of negotiation models for complex utility functions. Most of them have focused on mediated negotiation models. The seminal work of Klein et al. \cite{klein02} proposes a mediated negotiation model where agents' preferences are represented by influence matrices. Influence matrices represent binary interdependence relationships between binary issues. Their proposed approach consists of a mediator that generates bids that are voted by the agents participanting in the protocol. Ito et al. \cite{ito08} propose a mediated negotiation model for multilateral negotiations where agents have their preferences represented by weighted constraints. The agents sample their utility function and carry out a simulated annealing for each point sampled in order to obtain one's own bids. If the utility of such point is above a certain threshold, the constraints that the bid satisfies are sent to the mediator (constraint bid). After receiving bids from the agents, the mediator tries to look for contracts common to the bids received, while maximizing social welfare. Marsa-Maestre et al. \cite{marsa09,marsa09b} carry out further research in the area of mediated negotiation models for complex utility functions. More specifically, they take advantage of the constraint based model by proposing different bidding mechanisms that work in the constraint space instead of the bid space. They also allow for a negotiation protocol that may not be \textit{one-shot}. In fact, the mediator can suggest the relaxation of some constraint bids in order to increase the probability of finding an agreement. Nevertheless, all of these works need a trusted mediator, which may not be available in every domain. Furthermore, their models are highly dependent on the underlying utility function. The present work does not require a mediator and the model is independent of the underlying utility function.

Robu et al. \cite{robu05,robu06} presented a non-mediated bilateral negotiation strategy for agents in electronic commerce. Agent utility functions are based on special graphical models called utility graphs. One of the agents, the seller, is responsible for finding agreements that are satisfactory for both parties. In order to do that, the seller models the buyer by means of utility graphs and tries to learn the buyer's preferences. However, utility graphs are only designed for binary issues. Our work differs in that it is capable of working with general complex utility functions and is also capable of working issue domains that are not necessarily binary.

In Lai et al. \cite{lai08}, a powerful bilateral bargaining model with general utility functions is presented. The negotiation protocol is based on the Rubinstein alternating protocol \cite{rubi90}, but each agent is allowed to send up to \textit{k} different offers in each round. The offer with highest utility is chosen from the \textit{k} offers received from the opponent in the last round. The offer from the current iso-utility curve that is the most similar to the one chosen by the agent from the offers made by the opponent is selected. This offer from the iso-utility curve becomes a seed from which \textit{k-1} offers in the neighborhood are generated. The selected offer from the curve and the \textit{k-1} generated offers are sent back to the opponent. Again, the general ideal behind this heuristic is that, since the offers are similar to one of the last offers received from the opponent, they are more likely to be satisfactory for both parties. The model proposed in this article complements the seminal work of Lai et al. since it adapts similarity models for AmI environments. In the model proposed in this article, only a small number of offers are sampled before the negotiation process, since it is assumed that the utility function cannot be exhaustively explored. This is especially important for scenarios with a large number of issues and scenarios where devices have limited storage and computational resources. Secondly, the proposed model incorporates an implicit learning mechanism that allows, thanks to genetic operators, an intelligent sampling of new offers that may be of interest for both parties. 

\section{Conclusions and Future Work}
Ambient Intelligence aims to offer new services and methods of interaction with technology adapted to the users. It has been stated that automated negotiation may provide a conflict-resolution mechanism in Ambient Intelligence applications where several users with opposing preferences need to cooperate (e.g. ubiquitous shopping malls, fairs). In these environments, users' agents may present utility functions that are complex due to the interdependences among the negotiation issues that form the utility function.

A multi-issue bilateral bargaining model for Ambient Intelligence domains that deals with complex utility functions has been presented in this article. This work complements the inspiring work of Lai et al. \cite{lai08} and provides a negotiation model that is adequate for Ambient Intelligence applications. The main goal of this work is to achieve efficient agreements while maintaining the use of computational resources low. 

The proposed model uses a negotiation protocol where agents are allowed to send up to $k$ different offers in each negotiation round. Before the negotiation process starts, each agent samples its own utility function by means of a niching genetic algorithm. This genetic algorithm gets highly interesting and significantly different offers for one's own utility function (\textit{self-sampling}). After the negotiation process starts, the agents apply genetic operators over the last offers received from the opponent and those offers that are most similar from the current iso-utility curve (\textit{evolutionary sampling}). The desired effect is to sample new offers that are interesting for both parties. Therefore, the opponent preferences guide the sampling process during the negotiation process. The offers that are sent to the opponent are selected from the current iso-utility curve, being those that are the most similar to the last offers received from the opponent. An additional mechanism is introduced that allowing priority  to be given to those offers that come from the \textit{evolutionary sampling} iso-utility curve.

Several experimental scenarios have been carried out and studied. In these tests, the proposed model has been compared with a similarity heuristic that has access to all of the possible offers and a similarity heuristic that samples the same number of offers before the negotiation process by means of a niching genetic algorithm (\textit{NES}). The results show that the proposed model needs very few computational resources and storage to obtain statistically equivalent results to the ideal case where all of the offers are available. For instance, the full iso-utility curve consists of $10^6$ offers and the proposed model just samples 1510 offers in a negotiation setting where the number of issues is 6, and the number of offers sent per negotiation round is 3. Additionally, although the proposed model and the \textit{NES} model sample the same number of offers, the first one obtains better results. In fact, the \textit{NES} model needs to sample 3.64 times more offers to obtain similar results. The low computational cost and the efficient results make the proposed model very adequate for Ambient Intelligence domains.

Future work includes studying the effect of changing preferences during the negotiation process, (i.e., when the strategy is integrated with an argumentation mechanism), and introducing different agent behaviors (more self-interested, more cooperative, etc) by means of some modifications to genetic and selection operators. 

\section*{Acknowledgments}
This work is supported by TIN2008-04446, PROMETEO/2008/051, TIN2009-13839-C03-01, CSD2007-00022  of the Spanish government, and FPU grant AP2008-00600 awarded to V.Sánchez-Anguix.
\bibliographystyle{model1b-num-names}
\bibliography{memory}

\end{document}